\documentclass[fleqn,usenatbib]{mnras}

\usepackage{newtxtext,newtxmath}
\usepackage[T1]{fontenc}

\DeclareRobustCommand{\VAN}[3]{#2}
\let\VANthebibliography\thebibliography
\def\thebibliography{\DeclareRobustCommand{\VAN}[3]{##3}\VANthebibliography}

\usepackage{graphicx}	% Including figure files
\usepackage{amsmath}	% Advanced maths commands
\usepackage{float}[H]
\usepackage{aliascnt}
\usepackage{subfig}
\usepackage{multirow}
\usepackage{color}
\usepackage{verbatim}
\usepackage[table,xcdraw]{xcolor}
\newaliascnt{eqfloat}{equation}
\newfloat{eqfloat}{h}{eqflts}
\floatname{eqfloat}{Equation}

\title[\texttt{MargNet}: Faint \& Compact Source Classification]{Photometric identification of compact galaxies, stars and quasars using multiple neural networks}

\author[S. Chaini et. al.]{Siddharth Chaini,$^{1}$\thanks{E-mail: sidchaini@gmail.com}
Atharva Bagul,$^{1}$\thanks{E-mail: atharvabagul2000@gmail.com}
Anish Deshpande,$^{2}$
Rishi Gondkar,$^{3}$
Kaushal Sharma,$^{4}$
M. Vivek,$^{5}$
\newauthor Ajit Kembhavi$^{6}$
\\
$^{1}$Department of Physics, Indian Institute of Science Education and Research, Bhopal 462066, India\\
$^{2}$Department of Computer Science Engineering, Indian Institute of Technology, Bombay 400076, India\\
$^{3}$Department of Computer Science, Pune Institute of Computer Technology, Pune 411043, India\\
$^{4}$Millennium Institute of Astrophysics (MAS), Nuncio Monse{\~{n}}or S{\'{o}}tero Sanz 100, Providencia, Santiago, Chile\\
$^{5}$Indian Institute of Astrophysics, Koramangala, Bengaluru  560034, India\\
$^{6}$Inter University Centre for Astronomy and Astrophysics (IUCAA), Pune 411007, India
}

\date{Accepted XXX. Received YYY; in original form ZZZ}

\pubyear{2022}

\begin{document}
\label{firstpage}
\pagerange{\pageref{firstpage}--\pageref{lastpage}}
\maketitle

% Abstract of the paper
\begin{abstract}
We present \texttt{MargNet}, a deep learning-based classifier for identifying stars, quasars and compact galaxies using photometric parameters and images from the Sloan Digital Sky Survey (SDSS) Data Release 16 (DR16) catalogue. \texttt{MargNet} consists of a combination of Convolutional Neural Network (CNN) and Artificial Neural Network (ANN) architectures. Using a carefully curated dataset consisting of 240,000 compact objects and an additional 150,000 faint objects, the machine learns classification directly from the data, minimising the need for human intervention.
\texttt{MargNet} is the first classifier focusing exclusively on compact galaxies and performs better than other methods to classify compact galaxies from stars and quasars, even at fainter magnitudes. This model and feature engineering in such deep learning architectures will provide greater success in identifying objects in the ongoing and upcoming surveys, such as Dark Energy Survey (DES) and images from the Vera C. Rubin Observatory.
\end{abstract}

% Select between one and six entries from the list of approved keywords.
% Don't make up new ones.

%%%%%%%%%%%%%%%%%%%%%%%%%%%%%%%%%%%%%%%%%%%%%%%%%%

%%%%%%%%%%%%%%%%% BODY OF PAPER %%%%%%%%%%%%%%%%%%

\section{Introduction}
\label{sec:intro}
Large-scale surveys like the Sloan Digital Sky Survey \citep[SDSS; ][]{yorkSloanDigitalSky2000}, the Dark Energy Survey \citep[DES;][]{darkenergysurvey}, the Zwicky Transient Facility \citep[ZTF;][]{bellmZwickyTransientFacility2014}, and the Subaru Prime Focus Camera \citep{miyazakiSubaruPrimeFocus2002} collect photometric and spectroscopic data on millions of objects like stars, galaxies, and quasars. The forthcoming Vera C. Rubin Observatory and its Legacy Survey of Space and Time \citep[LSST;][]{ivezicLSSTScienceDrivers2019} will only add to the data deluge. By collecting approximately 15 terabytes every night \citep{ivezicLSSTScienceDrivers2019}, the LSST will make the manual analysis and classification of data impossible.

The use of machine learning in astronomy has  evolved rapidly over the last two decades \citep{baron2019machine}. While machine learning refers to any algorithm that learns a task by training on data, deep learning is a subset of machine learning in which representations are learned through multiple layers of neural networks. Development on this front has also benefited work in astronomy. Some examples of how machine learning and deep learning have aided astronomers are: star-galaxy classification \citep{philipDifferenceBoostingNeural2002, ballRobustMachineLearning2006, vasconcellosDecisionTreeClassifiers2010, abrahamPhotometricCatalogueQuasars2012, soumagnacStarGalaxySeparation2015, kimStarGalaxyClassification2017, clarkeIdentifyingGalaxiesQuasars2020}, stellar spectrum classification and interpolation \citep{kuntzerStellarClassificationSingleband2016, sharmaApplicationConvolutionalNeural2020, sharmaStellarSpectralInterpolation2020}, light curve classification \citep{lochnerPhotometricSupernovaClassification2016,mahabalMachineLearningZwicky2019,mollerSuperNNovaOpensourceFramework2020}, galaxy morphology classification \citep{dielemanRotationinvariantConvolutionalNeural2015,abrahamDetectionBarsGalaxies2018, dominguezsanchezImprovingGalaxyMorphologies2018, walmsleyGalaxyZooProbabilistic2020,barchiMachineDeepLearning2020,nair2022fraction}, photometric redshift estimation \citep{disantoPhotometricRedshiftEstimation2018,pasquetPhotometricRedshiftsSDSS2019a}, gravitational waves identification \citep{georgeDeepLearningRealtime2018} and gravitational lensing identification \citep{chengIdentifyingStrongLenses2020}. 
Our aim in this paper is to use deep learning to distinguish between stars, quasars and galaxies based on their images in the SDSS. In survey images, stars and quasars appear as point sources convolved with the point spread function (PSF), and the distinction between the two is conventionally made through their spectra, which requires large telescopes and is
time consuming even with multi-object spectrographs.  Machine learning enables the separation of stars from quasars using their colours alone with good accuracy \citep[see, for example][]{abrahamPhotometricCatalogueQuasars2012}
. A difficulty in this approach is that the colours of high redshift quasars appear different from those of low redshift quasars, which might intrinsically
have the same spectrum. Galaxies can be distinguished from stars (and quasars) on the basis of their morphology, as in the early work by \citet{sebok1986angular}, but the accuracy diminishes as the objects get fainter and more compact. A more reliable approach to classify in such cases is to use the difference in spectral energy distributions, either using spectra or using multiwavelength images. However, both these approaches are limited by the availability of large scale data.

As the availability of data has increased, machine learning has become a powerful tool for star-galaxy classification. Earlier methods used random forests \citep{vasconcellosDecisionTreeClassifiers2010} and support vector machines \citep{Fadely_Hogg_Willman_2012}, which are easier to train with relatively small amounts of data, but they have weak performance (80 - 90\% completeness). Newer methods involving convolutional neural networks \citep{kimStarGalaxyClassification2017, hao2017stacked, pmlr-v80-kennamer18a} perform significantly better, aided by the development of better image classification algorithms by the machine learning community for ImageNet 1 \citep{alexnet, szegedyGoingDeeperConvolutions2014, resnet, vcgnet}.

There is often a faint limit for magnitudes beyond which classification usually becomes difficult. For example, current methods degrade in performance when 
$r > 22.5$ \citep{cabayol2019pau} and $i > 21$
\citep{kimStarGalaxyClassification2017}. Newer sky surveys like the LSST will observe even deeper into the sky, making the importance of classification at faint magnitudes ever more important. We present a novel method to classify compact, faint sources with increased accuracy, in which we use a deep learning architecture that combines parametric and image-based classification.  Signal to noise ratio plays an important role in running our model on these newer datasets. It consists of a mixture (stacking ensemble) of a convolutional neural network (CNN) and an artificial neural network (ANN). Our training set consists of a set of photometric features and the images in five photometric bands (u, g, r, i, z) corresponding to the SDSS sources which have been spectroscopically classified as star, galaxy, or quasar by the SDSS pipeline. In order to select only the compact galaxies and fainter sources, we use some constraints on de Vaucouleur’s radius and photometric magnitude which are defined and explained in Section 2.   For the full application of \texttt{MargNet} we need the measured parameters (as mentioned in the Table \ref{tab:photoparam}) for the machine learning part. For data sets from other surveys, the measured parameters may not be identical to those for the SDSS. In such cases the machine learning network can be easily altered to accommodate the different set. If the parameters are not provided, then they could be measured by the user. If use of the parameters is not practical only the deep learning part could be used which needs just the images. \footnote{We have provided both pre-trained models on our \href{https://github.com/sidchaini/MargNet}{GitHub repository} mentioned in the data availability section.} We perform three experiments to look at different aspects of performance in the faint and compact limit by preparing three different sets of training, validation, and test samples. Many past studies focus only on the star-galaxy separation. 

Some earlier studies have also considered the full star galaxy and quasar separation problem \citep[e.g.:,][]{nakazono, Xiaoqing2020ClassificationOS}. To compare our results with those in recently published work, we separately perform star-galaxy classification as well as  star-galaxy-quasar classification in each of the three experiments. We show that our method performs better than the earlier methods for faint and compact galaxies , considering the limitation on the faintness and compactness parameters and subsequent limits on them as defined in the Section \ref{subsec:compact-faint}.

This paper is structured as follows: Section \ref{sec:data} covers data acquisition from SDSS DR16 Catalogue. Section \ref{sec:preprocessing} covers the preprocessing and Section \ref{sec:meth} the method employed for the classification problem along with details on the models. In Section \ref{sec:results} we provide the results and discussion and in Section \ref{sec:conclusion} the conclusions and further possibilities that arise from this work.

\section{Dataset}
\label{sec:data}

\subsection{SDSS Data}
\label{subsec:sdss-data}
Of the numerous sky surveys conducted over the past decades, few compare to the extent and detail of the Sloan Digital Sky Survey, or the SDSS;
\citep{yorkSloanDigitalSky2000}. The SDSS collected data with a dedicated 2.5 metre telescope \citep{gunnTelescopeSloanDigital2006}, CCD camera \citep{gunnSloanDigitalSky1998}, five filters \citep{doiPHOTOMETRICRESPONSEFUNCTIONS2010}, and a spectrograph \citep{smeeMULTIOBJECTFIBERFEDSPECTROGRAPHS2013}. The SDSS furnishes detailed, multicoloured images of approximately one-third of the sky and spectra for more than three million astronomical objects. For this paper, we make use of the SDSS Data Release 16 \citep{ahumadaSixteenthDataRelease2020}. 

We only use SDSS photometric data in the five passbands, namely u, g, r, i and z \citep{fukugitaSloanDigitalSky1996} as an input to our models. Since we train our models in a supervised fashion, an approach which requires a label corresponding to each input, we obtain the spectroscopically assigned class by the SDSS pipeline. 

We use two different types of data from the SDSS for our work: image data and photometrc features.

\begin{itemize}
    \item The image data in the five photometric filters, u g r i z, is extracted from the FITS (Flexible Image Transport System) files obtained in each filter from the SDSS.
    \item The photometric features \footnote{The definitions of all the parameters can be found at
    \href{http://skyserver.sdss.org/dr16/en/help/browser/browser.aspx?cmd=description+PhotoObjAll+U\#\&\&history=description+PhotoObjAll+U}{DR16 PhotObjAll}} listed in Table \ref{tab:photoparam} for each source are also used as input to the models. These features are obtained using the SDSS query service CasJobs. 
\end{itemize}

\begin{table}
\centering
\resizebox{\columnwidth}{!}{
\begin{tabular}{|c|c|c|} 
\hline
\textbf{Name} & \textbf{Description}                           & \textbf{Total Parameters}  \\ 
\hline
dered\_x      & Dereddened magnitude: corrected for extinction & 5                          \\
deVRad\_x     & DeVaucouleurs fit scale radius                 & 5                          \\
psffwhm\_x    & FWHM of the Point Spread Function              & 5                          \\
extinction\_x & Extinction                                     & 5                          \\
u\_g          & Colour: u-g                                    & 1                          \\
g\_r          & Colour: g-r                                    & 1                          \\
r\_i          & Colour: r-i                                    & 1                          \\
i\_z          & Colour: i-z                                    & 1                          \\ 
\hline
\multicolumn{2}{l|}{}                                          & 24                         \\
\cline{3-3}
\end{tabular}
}
\caption{ List of photometric parameters used. The colours given in the rows 5-8 are based on the dereddened magnitude. Note: Here x denotes one of the filters u/g/r/i/z.}

\label{tab:photoparam}
\end{table}

We use two steps to control the quality of our data.
First, we require that each object have an error in the PSF magnitude less than 0.2 across the r passband. 
Second, we use a combination of flags to limit our search to objects with clean photometry in our query. These flags\footnote{A detailed description of these flags is available on the  \href{https://www.sdss.org/dr16/tutorials/flags/\#SearchforObjectswithCleanPhotometry}{SDSS website}.} are:

\begin{itemize}
    \item \texttt{s.zWarning}=0 (no problem with spectra)
    \item \texttt{p.clean}=1 (clean photometry)
    \item \texttt{p.mode} = 1 (select primary objects only)
    \item \texttt{s.sciencePrimary} = 1 (no repeats in this query)
    \item \texttt{s.targetType} = 'SCIENCE'
    \item \texttt{p.insideMask}=0 (photometry without any masks)
    \item (\texttt{p.flags\_r} \& \texttt{0x20}) = 0 (not \texttt{PEAKCENTER})
    \item (\texttt{p.flags\_r} \& \texttt{0x80000}) = 0 (not \texttt{NOTCHECKED})
    \item (\texttt{p.flags\_r} \& \texttt{0x800000000000}) = 0 \\ \hspace*{0.6cm}(not \texttt{PSF\_FLUX\_INTERP})
    \item (\texttt{p.flags\_r} \& \texttt{0x10000000000}) = 0 \\ \hspace*{0.6cm}(not \texttt{BAD\_COUNTS\_ERROR})
    \item ((\texttt{p.flags\_r} \& \texttt{0x100000000000}) = 0 OR (p.flags\_r \& \\ \hspace*{0.6cm}0x1000) = 0) (not both \texttt{INTERP\_CENTER} and COSMIC\_RAY)
    \item (\texttt{p.flags\_r} \& \texttt{0x40000}) = 0 (not \texttt{SATURATED})
    \item (\texttt{p.flags\_r} \& \texttt{0x80}) = 0 (not \texttt{NOPROFILE})
\end{itemize}

\subsection{Compactness \& Faintness}
\label{subsec:compact-faint}

\begin{figure}
    \centering
    \includegraphics[width = \columnwidth]{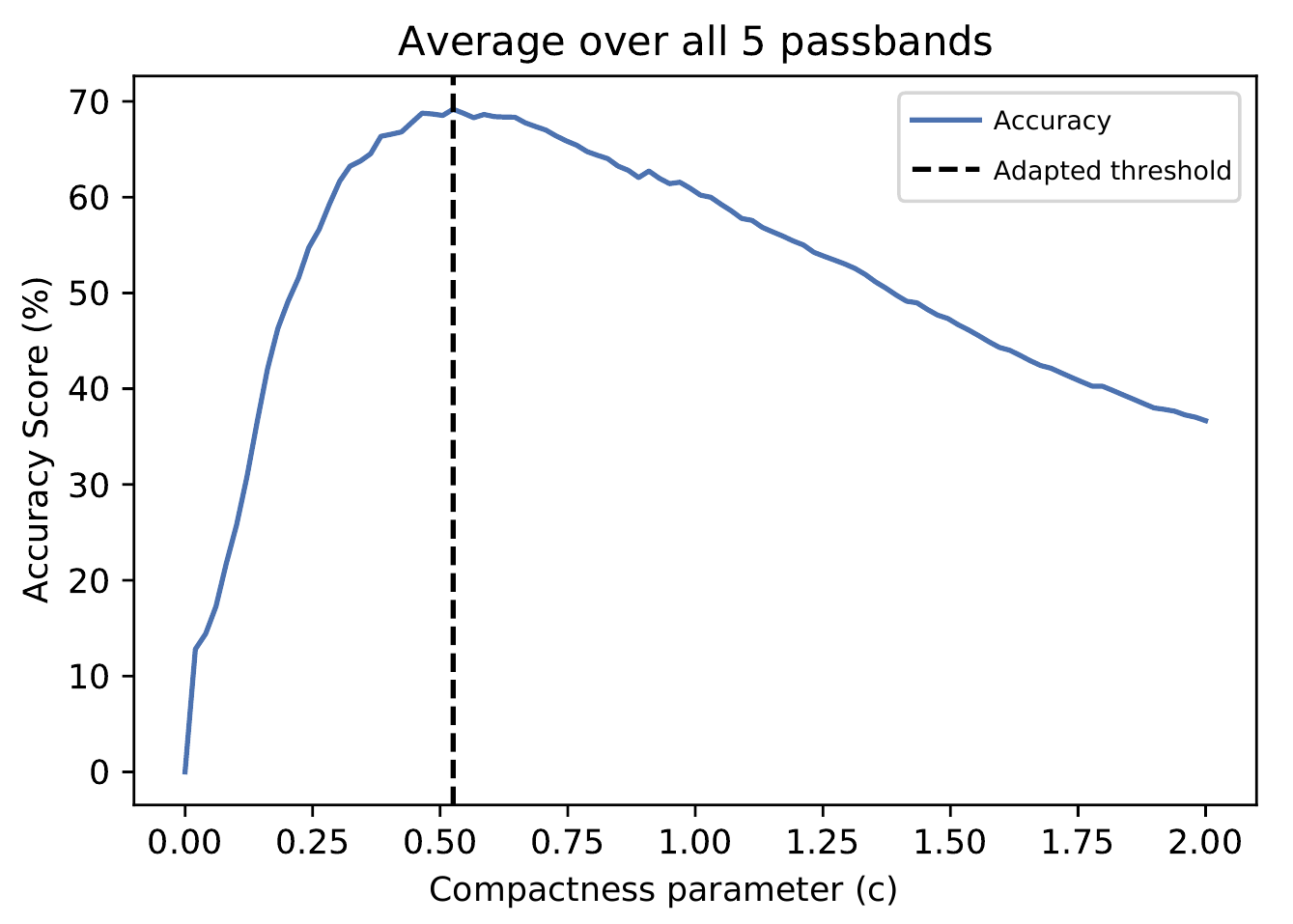}
    \caption{Model performance vs the compactness parameter, $c = \langle \dfrac{\mathrm{deVRad}}{\mathrm{FWHM}} \rangle$}
    \label{fig:marginality}
\end{figure}

As described in Section \ref{sec:intro}, we aim to identify galaxies which are too compact to properly separate from stars and quasars using traditional methods. Astronomers have used different ways of measuring the compactness of sources, which lead to nearly the same results. While some definitions of compactness are based on the galaxy's magnitude \citep{zwickycompact1978, hicksonSystematicPropertiesCompact1982}, we do not use any magnitude criterion in our work, because low surface brightness (LSB) galaxies can have extended morphology albeit being faint. Instead, we define the compactness parameter $c$ for an object as:
\begin{equation} \label{eq:compactness}
    \hspace{3.4cm}c = \Big\langle \frac{\mathrm{deVRad}}{\mathrm{FWHM}} \Big\rangle
\end{equation}
Here deVRad is the half light radius (also known as de Vaucouleur's radius) of the galaxy. $\langle \rangle$ denotes an average over all the 5 passbands: u, g, r, i and z. FWHM denotes the full width at half maximum of the point spread function (PSF). For a given FWHM, smaller the value of $c$, the more compact is the galaxy. 
To establish the limit on $c$ below which classification becomes difficult, we trained a Random Forest classifier using photometric parameters (See Section \ref{sec:data} for details) on a small random dataset of spectroscopically identified stars, galaxies, and quasars, and evaluated the performance using the accuracy achieved in the classification. By retraining the Random Forest for training samples in small bins of $c$, we can see how the value of $c$ affects the performance, as shown in Fig  \ref{fig:marginality}. We observe a sharp drop in the performance when $c$ becomes $<0.5$. We therefore use $c = 0.5$ as the boundary below which we classify galaxies as compact. The drop in performance for values of $c > 0.5$ is due to the fewer number of galaxies available for training the Random Forest classifier.
This finding qualitatively implies that when $c$ is less than or equal to 0.5 (the galaxy’s diameter is roughly equal to the FWHM of its PSF), the automated classification becomes challenging due to the compactness of the galaxy.

When choosing objects to constitute our faint dataset, we require that the average magnitude in the 5 passbands
%\vspace{-10pt}
\begin{equation}
    \label{eq:faint}
    \hspace{3.5cm}\langle \textrm{mag} \rangle > 20
\end{equation}
We choose $20$ as the cut-off point for two reasons. Firstly, it is beyond this level of faintness that the traditional star-galaxy classifiers start to fail, with a huge decrease in performance at $r>21$ \citep{kimStarGalaxyClassification2017, cabayol2019pau}. Secondly, enough samples from the SDSS obey this cut-off. A neural network needs sufficient amount of data to train properly, and a higher cut-off would have resulted in too small a dataset.

\subsection{Querying Faint and Compact Sources}

As described in Section \ref{subsec:compact-faint} we create two sets of data for our three experiments:
\begin{enumerate}
\item \noindent \hspace{2.25pt}Compact Source Dataset
\item \noindent Faint and Compact Source Dataset
\end{enumerate}
Because classification algorithms trained on an imbalanced dataset perform poorly on minority classes, we prefer to work with a balanced number of classes throughout this paper.  

\subsubsection{Compact Source Dataset}
\label{subsub:compact}
We impose the condition $c<0.5$ from Eq. \ref{eq:compactness} for querying the marginal, compact galaxies. While the de Vaucouleur's radius is generally only applied to galaxies, we apply our condition to all samples here in order to maintain uniformity in our data.

In DR16, there are 1,73,813 stars, 80,779 galaxies, and 527,922 quasars which are spectroscopically identified and meet all of the above criteria for the compact source dataset. We choose a random sample of 80,000 objects of each class to form our compact source dataset. 

\subsubsection{Faint and Compact Source Dataset}
\label{subsub:compactfaint}
In addition to the compact criterion $c<0.5$ from Eq. \ref{eq:compactness}, here we also require the condition $\langle mag \rangle > 20$ from Eq. \ref{eq:faint}. This gives us a dataset which contains sources that are faint as well as compact.

In DR16, there are 55,323 stars, 74,503 galaxies, and 382,899 quasars that meet all of the above criteria and we choose a random sample of 50,000 objects of each class to form our faint and compact source dataset. After querying the photometrc features and downloading the FITS images for both datasets, we are ready to run the three experiments.

\section{Preprocessing}
\label{sec:preprocessing}
\subsection{Image Data}
\begin{figure}
\centering
\includegraphics[width=\columnwidth]{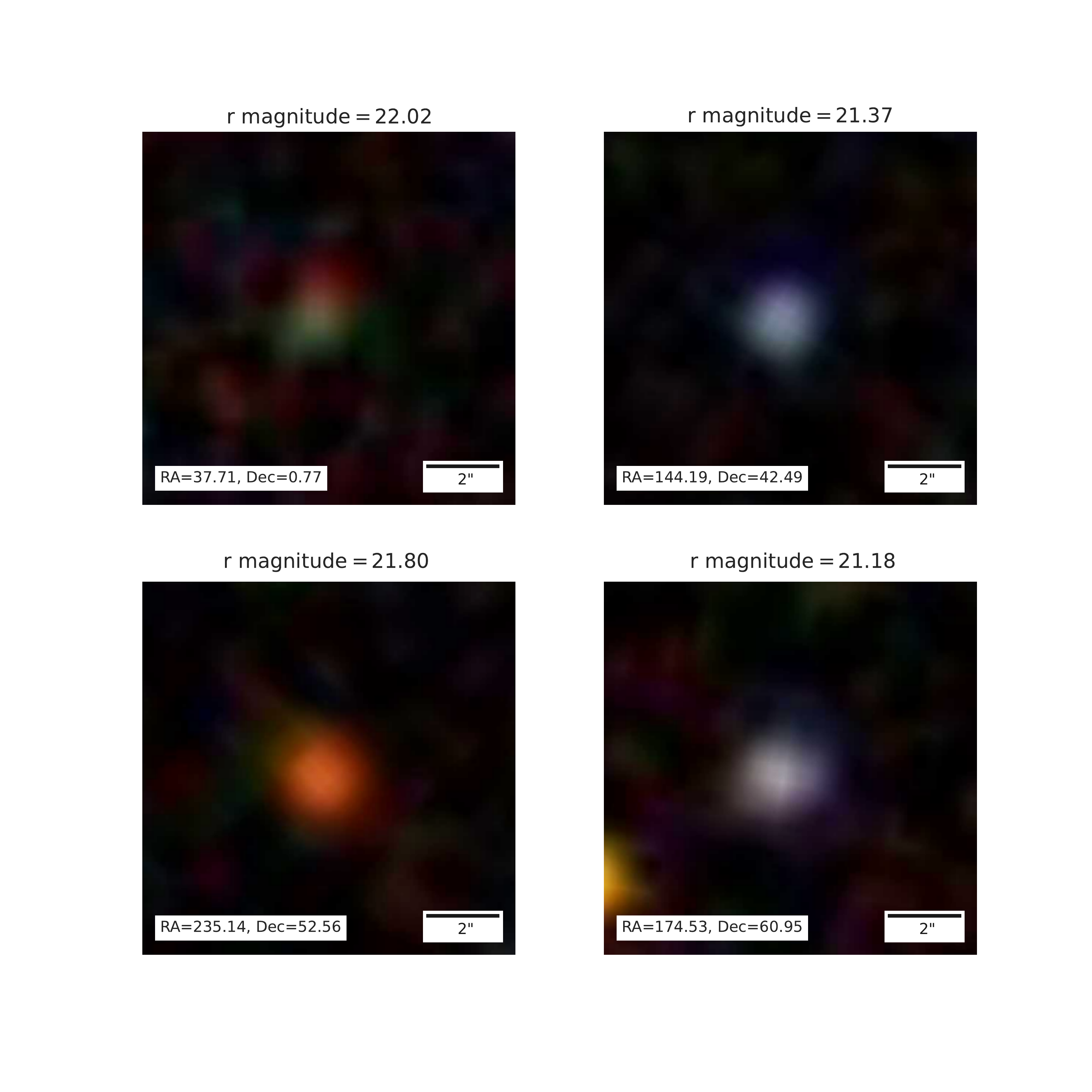}
\caption{A sample of 4 compact galaxies from our dataset. The images are constructed from the g-r, r-i and i-z colours obtained from the passbands.}
\label{fig:marg_samples}
\end{figure}

We use FITS images in the 5 passbands. However, there are a couple of issues that need to be solved during preprocessing.
We crop each FITS image to a $32 \times 32$ pixel image to remove any undesired features or nearby sources from the input image. We find these dimensions sufficient for our compact objects of interest. The right ascension (RA) and declination (Dec) values of the object of interest specify its pixel location in the image.  This is used as a reference point for the image ‘centre’. The image is cropped using the World Coordinate Systems on \texttt{Astropy} in Python. We specify a $32 \times 32$ pixel bounding box aligned with the axes around the centre for this purpose. Since each passband has a separate cropped image, we  stack them on top of each other. We thus produce a 5-channel image (Figure \ref{fig:marg_samples}) of dimension $32 \times 32 \times 5$ for each of the 240,000 objects.

\subsection{Photometric features}
The photometric parameters, as given in Table \ref{tab:photoparam}, have different ranges. Normalisation of a dataset is a common requirement for many machine learning algorithms. Thus, we normalise the data by removing the mean and scaling to unit variance for every individual parameter using \texttt{StandardScaler} \footnote{The documentation of StandardScaler function from SciKit-Learn can be found \href{https://scikit-learn.org/stable/modules/generated/sklearn.preprocessing.StandardScaler.html}{here}.} from the \texttt{scikit-learn} \citep{scikit-learn} library on Python.

\subsection{Train-Validation-Test Split}
As is common in deep learning, we split our dataset into a training set, validation set and a test set. In each of the three experiments, we select an approximately equal number of classes. Each subset is chosen differently depending on the experiment, full details of which are given in Table \ref{tab:distribution} and Figure \ref{fig:distribution}.

\subsubsection{Experiment 1}
In this experiment, all three sets - training, validation and test, are chosen from the Compact Source dataset (Section \ref{subsub:compact}: $c<0.5$), which is split in the ratio 6:1:1  (i.e 75\% training, 12.5\% validation and 12.5\% test). We note that the test set is representative of the training set here.

\subsubsection{Experiment 2}
In this experiment, all three sets - training, validation and test, are chosen from the Faint and Compact Source dataset (Section \ref{subsub:compactfaint}: $c<0.5; \langle mag \rangle > 20$) and split in the ratio 8:1:1  (i.e 80\% training, 10\% validation and 10\% test).
Here, we split our data differently than Experiment 1 and Experiment 3 as there were just 50,000 objects from each class in our dataset (as described in Section \ref{subsub:compactfaint}). So to ensure that our training set had enough training data, we changed our data split so that at least $40,000$ objects of each class were used for training in each experiment. This resulted in us using the ratio 8:1:1 (i.e 80\% training, 10\% validation and 10\% test).
We note that the test set is representative of the training set here.

\subsubsection{Experiment 3}
In this experiment, the training and validation set is chosen from the Compact Source dataset (Section \ref{subsub:compact}: $c<0.5$). However, the test set is chosen from the Faint and Compact Source dataset (Section \ref{subsub:compactfaint}: $c<0.5; \langle mag \rangle > 20$), such that the ratio is 6:1:1  (i.e 75\% training, 12.5\% validation, 12.5\% test). We also note that the test set is \textbf{not} representative of the training set here.

This experiment aims to see how well we can extend the training at brighter levels to fainter levels. This case will be significant when surveys like the LSST are in operation. This experiment aims to see how well we can extend the training at brighter levels to fainter levels. This case will be significant when surveys like the LSST are in operation. The spectroscopic dataset containing both faint and compact sources will have a relatively small number of objects and will pose a challenge for training a photometric classifier. Instead, we wish to look at how a model trained only on compact sources tests on faint and compact sources.

\begin{figure}
\centering
\includegraphics[width=\columnwidth]{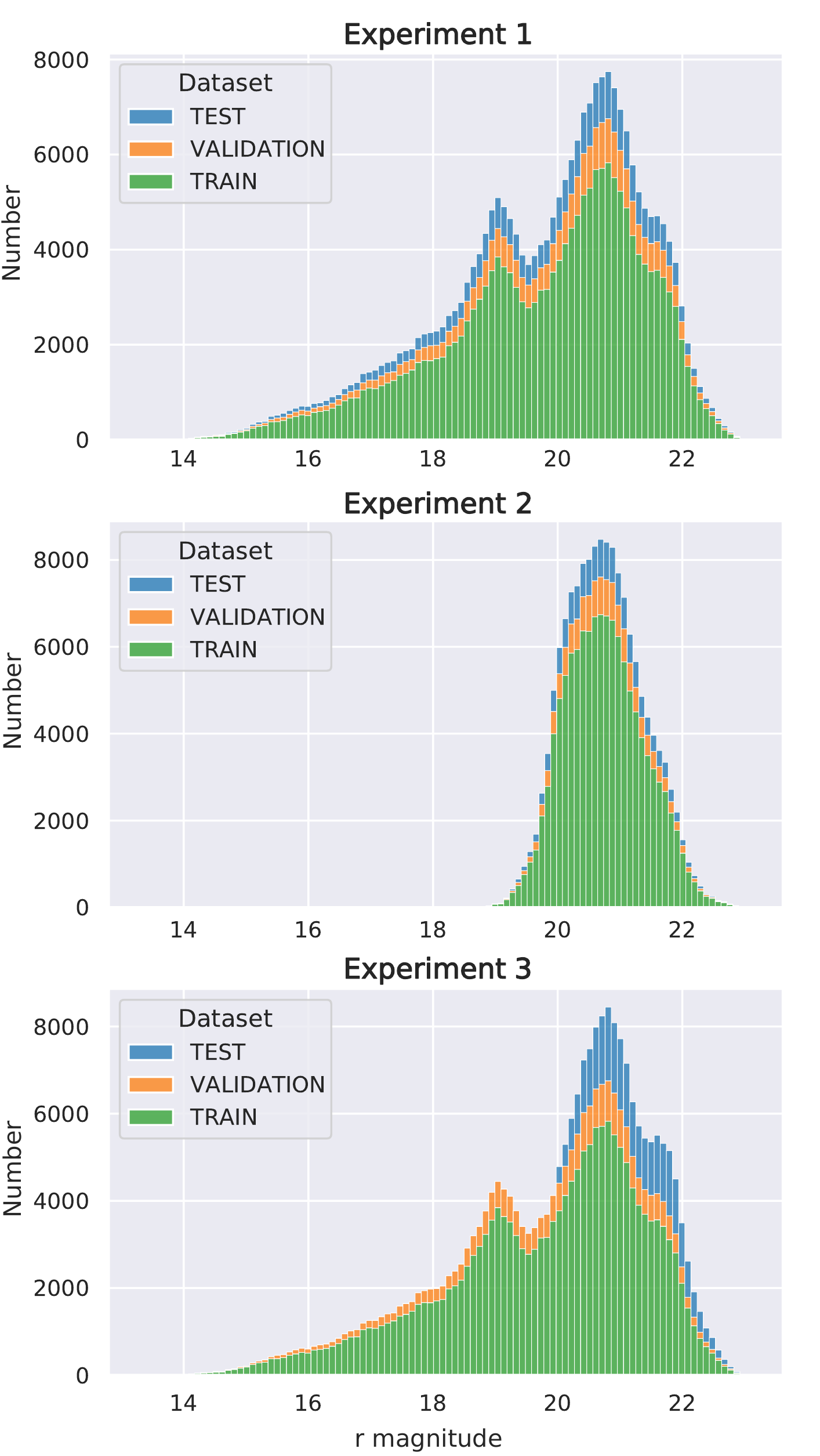}
\caption{Number distribution of our dataset as a function of magnitude in the $r$ band for each experiment.  Each image is a bar graph with each bar having Train, Validation and Test data in different colours being stacked above one another.
}
\label{fig:distribution}
\end{figure}

\begin{table*}
\centering
\begin{tabular}{|c|c|c|ccc|c|} 
\hline
\textbf{Experiment No.}       & \textbf{Dataset}                & \textbf{Subset} & \textbf{\# Stars} & \textbf{\# Galaxies} & \textbf{\# Quasars} & \textbf{Total}                              \\ 
\hline
\multirow{3}{*}{Experiment 1} & \multirow{3}{*}{Compact}        & Train           & 59988             & 60061                & 59962               & 180011                                      \\
                              &                                 & Validation      & 10033             & 9918                 & 10037               & 29988                                       \\
                              &                                 & Test            & 9979              & 10020                & 10001               & 30000                                       \\ 
\hline
\multirow{3}{*}{Experiment 2} & \multirow{3}{*}{Faint \&  Compact} & Train           & 39870             & 40134                & 39992               & \textcolor[rgb]{0.125,0.129,0.141}{119996}  \\
                              &                                 & Validation      & 5095              & 4956                 & 4949                & \textcolor[rgb]{0.125,0.129,0.141}{15000}   \\
                              &                                 & Test            & 5031              & 4910                 & 5059                & \textcolor[rgb]{0.125,0.129,0.141}{15000}   \\ 
\hline
\multirow{3}{*}{Experiment 3} & \multirow{2}{*}{Compact}        & Train           & 59988             & 60061                & 59962               & \textcolor[rgb]{0.125,0.129,0.141}{180011}  \\
                              &                                 & Validation      & 10033             & 9918                 & 10037               & \textcolor[rgb]{0.125,0.129,0.141}{29988}   \\ 
\cline{2-2}
                              & Faint \& Compact                  & Test            & 9511              & 9511                 & 9510                & 28532                                       \\
\hline
\end{tabular}
\caption{The distribution of our dataset for each experiment. We use an approximately equal of number of classes in each subset. In Experiment 1, all the sets consist of compact sources $(c<0.5)$, while in Experiment 2, all the sets consist of faint and compact sources $(c<0.5; \langle mag \rangle > 20)$. But, in Experiment 3, the training and validation set consist of compact sources $(c<0.5)$, while the test set consists of faint and compact sources $(c<0.5; \langle mag \rangle > 20)$}
\label{tab:distribution}
\end{table*}

\section{Methodology}
\label{sec:meth}

The approach consists of two parallel but separate machine learning models pipelined to create a functioning classifier. The first unit is a Convolutional Neural Network (CNN) \citep{lecun2015deep} model that takes the preprocessed images (processed as per Sec \ref{sec:preprocessing}) as the input. The second unit is an Artificial Neural Network (ANN) model \citep{mcculloch1943logical}, which takes the photometric parameters as input. These two units individually make an initial classification. The ensemble takes the initial classifications of CNN and ANN units as inputs and makes the final Star, Galaxy, Quasar classification.

\subsection{ANN Architecture}

\begin{figure}
\centering
\includegraphics[width=\columnwidth]{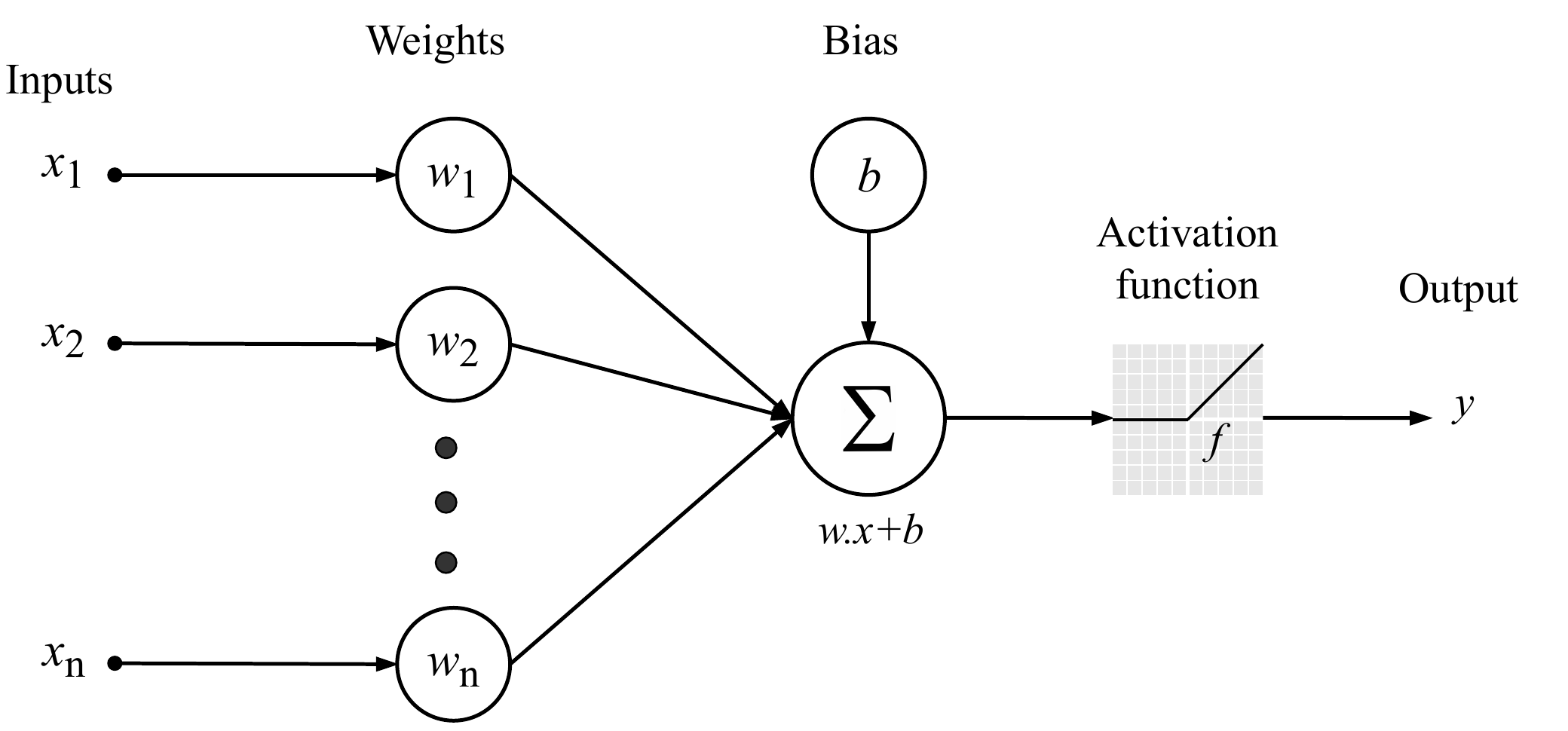}
\caption{Schematic representation of an artificial neuron. Inputs $(x_1, x_2,...,x_n)$ are multiplied by their corresponding weights $(w_1, w_2,...,w_n)$.  and a bias is added to the weighted inputs. An activation function is applied to the resultant which provides the final output $y$.}\label{fig:neural_net}
\end{figure}

 Artificial Neural Networks draw inspiration from Biological Neural Networks consisting of biological neurons as basic units. A biological neuron receives input signals through dendrites, processes the received information, and provides the output. Similarly, an artificial neural network consists of artificial neurons where each neuron receives the input through input nodes, processes the information, and provides an output. Mathematically, the output $y$ of an artificial neuron can be represented as follows:
\begin{equation}
    y = f\left(\sum w\cdot x + b \right)
\end{equation}
where $x$ is a vector of inputs, $(x_1, x_2,...,x_n)$, and $w$ is a vector of weights $(w_1, w_2,...,w_n)$ corresponding to each input. A bias value $b$ is added to the summed input $w\cdot x$ which is further acted upon by a function $f$, referred to as the activation function. A schematic representation of an artificial neuron is shown in Figure \ref{fig:neural_net}. A practical ANN consists of more than one such neuron arranged in multiple layers having interconnections between neurons in different layers. The training of the neural network refers to the learning of the weights $w$ and bias value(s) $b$ in order to minimize the departure between the neural output $y$ and the expected output, say $\bar{y}$. A widely used non-linear activation function is the rectified linear unit \citep[ReLU;][]{nair2010rectified}. Since ReLU units (Eq. \ref{eq:relu}), in general, allow faster training of deep neural networks with many layers, it was our natural choice for activation function in this exercise.
\begin{equation}\label{eq:relu}
    \sigma (x) = \begin{cases}x & x \geq 0\\0.01x & x < 0\end{cases}
\end{equation}

\subsection{CNN Architecture}
A Convolutional Neural Network (CNN) is a type of feed-forward neural network. These networks have gained significant use in computer vision applications like classification, counting, and a plethora of others. 
The main difference between a regular  neural network (ANN) and a CNN is that the latter has  convolution operations between a set of filters  and the inputs. A convolution  is expressed as
\begin{equation}
    y^k = \sigma\left(\sum\limits_{m} w^k_m * x_m + b^k \right)
\end{equation}
where we sum over the set of input feature maps, "$*$" is the convolution operator, and $w$ represents the filter weights. Here, a feature map is the array of output activation obtained after applying the activation function.
In a typical CNN, there are three kinds of layers: convolution layers, pooling layers, and fully connected layers. A pooling layer is present after a convolutional layer; it is used to to perform down-sampling of the feature map by way of a 'pooling' operation. This could, for example, lead to  the average value of $n\times n$ pixel values, where $n$ is chosen typically to be a small value like 2 or 3. This is done for all patches in the image, with stride length $n$ and is carried forward as the output of an average-pooling layer, so that the image is down-sampled by a factor of $n$ in each dimension. The primary function of the pooling layers is to reduce the dimension of the feature map and make the model invariant to small shifts and distortions. Similarly, max-pooling involves calculating the maximum value in a patch and using that as the layer's output. The fully connected layers are used at the end of a CNN which take the output of the convolutional and pooling layers to predict the best label that describes the image. The layers in CNNs are sparser than fully connected layers. In sparse connections, a neuron in one layer receives input from a local patch of neurons in the previous layer, whereas in fully connected layers the input received by a neuron is from every possible neuron in the previous layer. This sparsity reduces the number of connections in the CNN. Also, the kernel that is used for the convolution of a layer uses the same weights across the whole layer. This weight sharing greatly reduces the number of unique parameters in the network making CNNs relatively easier to train than fully connected networks.

A CNN takes in the pre-processed image files for all three data objects to provide an initial classification as the output. We have 80,000 images for each class with the available spectroscopic labels. The CNN uses only the images, without parameters. Each image is processed through a deep network of convolutional layers, of varying kernel sizes (kernels used: [1,1], [3,3], [5,5]), each activated by a ReLU unit \citep{nair2010rectified}.
A dense layer is one where each neuron in the layer receives an input from every single neuron in the previous layer. Every possible pair has a weighted connection between its members. The information is distilled through such final few dense layers, and the final layer contains a softmax (generalisation of the logistic regression function used for multi-class classification) function to divide each input as a probability distribution across the three classes. 
We employ a frequently used a widely-used loss function, categorical cross-entropy.
The CNN model trains a total of 25,544,807 neurons. We have used a CNN architecture which is inspired by the Inception Network created by \citet{szegedyGoingDeeperConvolutions2014} for the ImageNet Challenge. InceptionNet has been used successfully in astronomy, and we employ the layer sizes used by \citet{pasquetPhotometricRedshiftsSDSS2019a} for photometric redshift estimation. 
The inception modules are useful because they help extract the most discriminating features from different sizes of convolution filters. So filters of size (1x1), (3x3) and (5x5) are applied at one layer of the network, on whatever the input is at that depth. Each output is concatenated to form one thick layer. We perform this operation every few layers, and each such operation defines our 'inception module'. It is up to the network to weigh different filter outputs during training.
The implementation details for this are available in Table \ref{tab:CNNmodel}.

We do not use dropout fraction (details for dropout fraction are given in Sec \ref{subsec:dropout}) for CNN as it is used for densely connected layers, not in layers with sparse connections like the convolutional layers in the CNN. Dropout fraction is applied in neural networks which have significantly fully-connected components (e.g.: ANN). This prevents the large number of parameters present from overfitting the data.

\begin{table*}
\centering
\begin{tabular}{|c|c|c|c|c|}
\hline
\textbf{Layer} & \textbf{Input}          & \textbf{\# Filters} & \textbf{Kernel Size} & \textbf{Activation} \\ \hline
Conv2D\_1             & input image             & 64                  & (5,5)                & ReLU                \\ \hline
Conv2D\_2             & Conv2D\_1                      & 48                  & (1,1)                & ReLU                \\ \hline
Conv2D\_3             & Conv2D\_1                      & 48                  & (1,1)                & ReLU                \\
Conv2D\_4             & Conv2D\_1                      & 48                  & (1,1)                & ReLU                \\
AvgPool2D\_1             & Conv2D\_4                      & -                   & -                    & -                   \\
Conv2D\_5             & Conv2D\_2                      & 64                  & (1,1)                & ReLU                \\
Conv2D\_6             & Conv2D\_2                      & 64                  & (3,3)                & ReLU                \\
Conv2D\_7             & Conv2D\_3                      & 64                  & (5,5)                & ReLU                \\
Concatenate\_1            & {[}Conv2D\_5, Conv2D\_6. Conv2D\_7, AvgPool2D\_1{]}    & -                   & -                    & -                   \\ \hline
Conv2D\_8             & Concatenate\_1                     & 64                  & (1,1)                & ReLU                \\
Conv2D\_9             & Concatenate\_1                     & 64                  & (1,1)                & ReLU                \\
Conv2D\_10            & Concatenate\_1                     & 64                  & (1,1)                & ReLU                \\
AvgPool2D\_2             & Conv2D\_10                     & -                   & -                    & -                   \\
Conv2D\_11            & Conv2D\_8                      & 92                  & (1,1)                & ReLU                \\
Conv2D\_12            & Conv2D\_8                      & 92                  & (3,3)                & ReLU                \\
Conv2D\_13            & Conv2D\_9                      & 92                  & (5,5)                & ReLU                \\
Concatenate\_2            & {[}Conv2D\_11, Conv2D\_12, Conv2D\_13, AvgPool2D\_2{]} & -                   & -                    & -                   \\ \hline
AvgPool2D\_3             & Concatenate\_2                     & -                   & -                    & -                   \\ \hline
Conv2D\_14            & AvgPool2D\_3                      & 92                  & (1,1)                & ReLU                \\
Conv2D\_15            & AvgPool2D\_3                      & 92                  & (1,1)                & ReLU                \\
Conv2D\_16            & AvgPool2D\_3                      & 92                  & (1,1)                & ReLU                \\
AvgPool2D\_4             & Conv2D\_16                     & -                   & -                    & -                   \\
Conv2D\_17            & Conv2D\_14                     & 128                 & (1,1)                & ReLU                \\
Conv2D\_18            & Conv2D\_14                     & 128                 & (3,3)                & ReLU                \\
Conv2D\_19            & Conv2D\_15                     & 128                 & (5,5)                & ReLU                \\
Concatenate\_3            & {[}Conv2D\_17, Conv2D\_18, Conv2D\_19, AvgPool2D\_4{]} & -                   & -                    & -                   \\ \hline
Conv2D\_20            & Concatenate\_3                     & 92                  & (1,1)                & ReLU                \\
Conv2D\_21            & Concatenate\_3                     & 92                  & (1,1)                & ReLU                \\
Conv2D\_22            & Concatenate\_3                     & 92                  & (1,1)                & ReLU                \\
AvgPool2D\_5             & Conv2D\_22                     & -                   & -                    & -                   \\
Conv2D\_23            & Conv2D\_20                     & 128                 & (1,1)                & ReLU                \\
Conv2D\_24            & Conv2D\_20                     & 128                 & (3,3)                & ReLU                \\
Conv2D\_25            & Conv2D\_21                     & 128                 & (5,5)                & ReLU                \\
Concatenate\_4            & {[}Conv2D\_23, Conv2D\_24, Conv2D\_25, AvgPool2D\_5{]} & -                   & -                    & -                   \\ \hline
AvgPool2D\_6             & Concatenate\_4                     & -                   & -                    & -                   \\ \hline
Conv2D\_26            & AvgPool2D\_6                      & 92                  & (1,1)                & ReLU                \\
Conv2D\_27            & AvgPool2D\_6                      & 92                  & (1,1)                & ReLU                \\
AvgPool2D\_7             & Conv2D\_27                     & -                   & -                    & -                   \\
Conv2D\_28            & AvgPool2D\_6                      & 128                 & (1,1)                & ReLU                \\
Conv2D\_29            & Conv2D\_26                     & 128                 & (3,3)                & ReLU                \\
Concatenate\_5            & {[}Conv2D\_29, AvgPool2D\_7, Conv2D\_28{]}      & -                   & -                    & -                   \\ \hline
Dense\_1             & Concatenate\_5                     & 1024*               &                      & None                \\ \hline
Dense\_2             & Dense\_1                      & 1024*               & -                    & None                \\ \hline
Dense\_3             & Dense\_2                      & 3*                  & -                    & softmax             \\ \hline
\end{tabular}
\caption{Characteristics of each layer of the CNN architecture: Column 1: name of the layer, Column 2: input layer, Column 3: number of filters in the convolution kernel, Column 4: size of the convolution kernel (in pixels) and Column 5: activation function used for the layer.}
\label{tab:CNNmodel}.
\end{table*}

\subsection{Artificial Neural Network}
\label{subsec:ann}

\begin{figure}
\centering
\includegraphics[height=8cm]{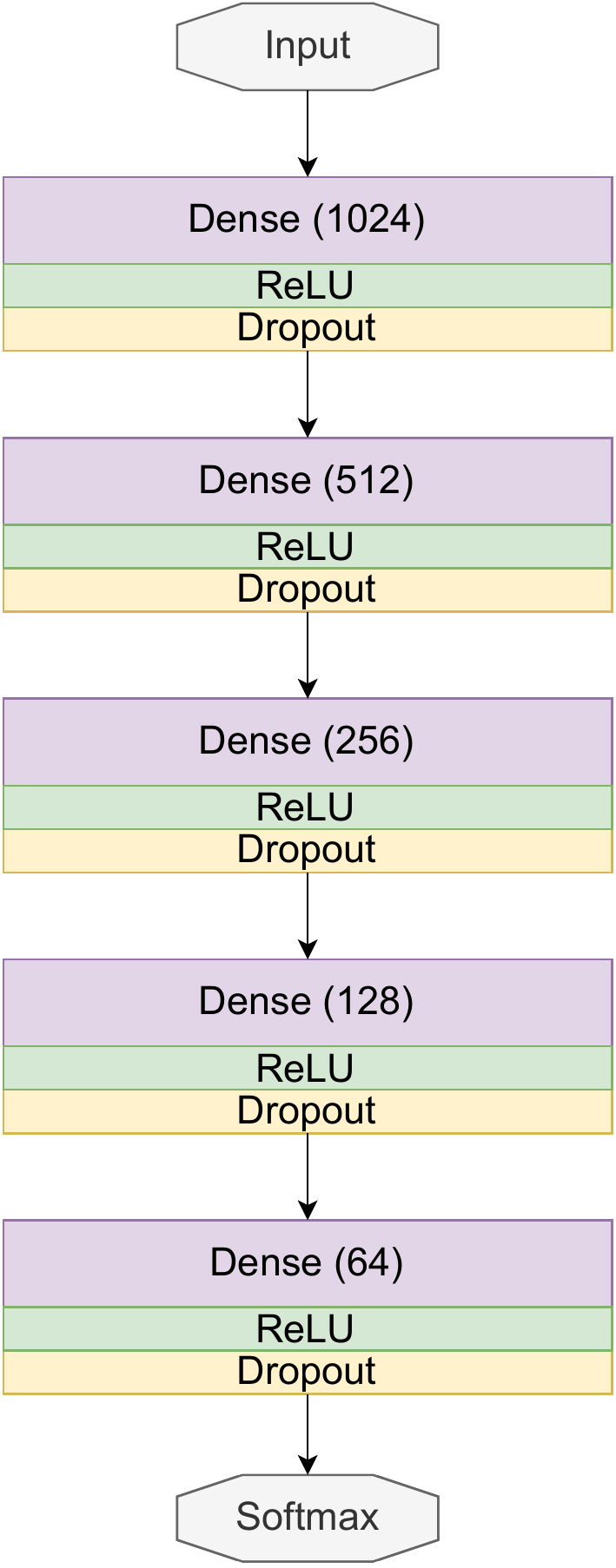}
\caption{Artificial Neural Network Architecture: This contains five dense layers of size 1024, 512, 256, 128, 64 respectively, each activated with ReLU, along with a dropout fraction of 0.25. It trains a total of 331,331 neurons. For more details refer to Section \ref{subsec:ann}.}
\label{fig:DNN}
\end{figure}

In this component, the photometric parameters of the images are used for input. The images are not considered, and hence no additional preprocessing (other than normalisation) is required for this component. It is a feed-forward network with stacked dense layers, each activated by a ReLU unit \citep{nair2010rectified}. The five dense layers ensure enough depth for direct classification from the selected list of parameters. The presence of dropouts (refer to the Section \ref{subsec:dropout}) (dropout fraction = 0.25) ensures overfitting is taken care of, which is a problem, especially in deep networks (see next section). The final layer is a softmax-activated with three outputs, one for each category. Again,  categorical cross-entropy is a natural choice for the loss function. The model trains a total of 331,331 neurons and the implementation details are available in Table \ref{tab:annmodel}. Table \ref{tab:CNNmodel} and Table \ref{tab:annmodel} combined simultaneously can be used to implement \texttt{MargNet}, as given in Figure \ref{fig:ensemble_model}.

\begin{table*}
\centering
\begin{tabular}{|c|c|c|c|c|}
\hline
\textbf{Layer} & \textbf{Input} & \textbf{Neurons} & \textbf{Activation} & \textbf{Dropout} \\
\hline
Dense\_4 & photometric features & 1024 & ReLU & 0.25 \\
Dense\_5 & Dense\_4 & 256 & ReLU & 0.25 \\
Dense\_6 & Dense\_5 & 128 & - & 0.25 \\
Dense\_7 & Dense\_6 & 32 & ReLU & 0.25 \\
Dense\_8 & Dense\_7 & 64 & ReLU & 0.25 \\
Dense\_9 & Dense\_8 & 3 & softmax & 0 \\
\hline
\end{tabular}
\caption{Characteristics of each layer of the ANN architecture: Column 1: name of the layer, Column 2: input layer, 
Column 3: refers to the number of neurons in the layer, Column 4: activation function used for the layer and Column 5: value of the dropout function for each layer in the ANN.}
\label{tab:annmodel}
\end{table*}

\subsection{Dropout \& Image Augmentation}
\label{subsec:dropout}
When neural networks are trained on small or homogeneous datasets, the models tend to overfit the data \citep{JMLR:v15:srivastava14a}. They do not generalise effectively to perform well on the validation and test datasets. To mitigate overfitting, a technique we employ is the inclusion of dropout in the network \citep{SrivastavaDropout}. A fraction (called the dropout fraction) of  neurons are randomly chosen and artificially set to 0 only during training. This helps prevent overfitting of the neural network. 

Data augmentation is another effective method that can be used to minimise overfitting. We create variations in the data, which are fed to the model while training. This artificially created and enhanced version of the data improves model learning, enabling the model to generalise well during the testing phase \citep{perez2017effectiveness}. We augment the procured data points using Python Keras tools, namely the \texttt{ImageDataGenerator} module \footnote{Details and documentation of the \texttt{ImageDataGenerator} module can be found} \href{https://www.tensorflow.org/api_docs/python/tf/keras/preprocessing/image/ImageDataGenerator}{here}, to make our model more robust. 
We apply the following transforms independently and randomly over the dataset:
(1) We mirror images randomly along the horizontal or the vertical axis; 
(2) We rotate images by an angle between 1 and 180 degrees.
The image background is mostly monochromatic and of little use; hence the cropping of edges during rotation is neglected; 
(3) We translate images along the vertical and horizontal axis by 1 to 3 pixels.
The resultant images are affine transformations of the originals, defined by 
\begin{equation}
    y=W x+b
\end{equation}
Here $W$ is the \textit{affine transformation matrix} used, $y$ is the affine transformation map, and $b$ is a bias vector.
%constant parameter
\citep{weisstein2004affine}.
%%AB: Look b up.
Affine transforms maintain parallelism and lines, and hence do not destroy or fabricate astronomical information in the images relevant to us.

\subsection{Ensemble (\texttt{MargNet})}
\label{subsec:marg}
\begin{figure}
\centering
\includegraphics[height=3cm]{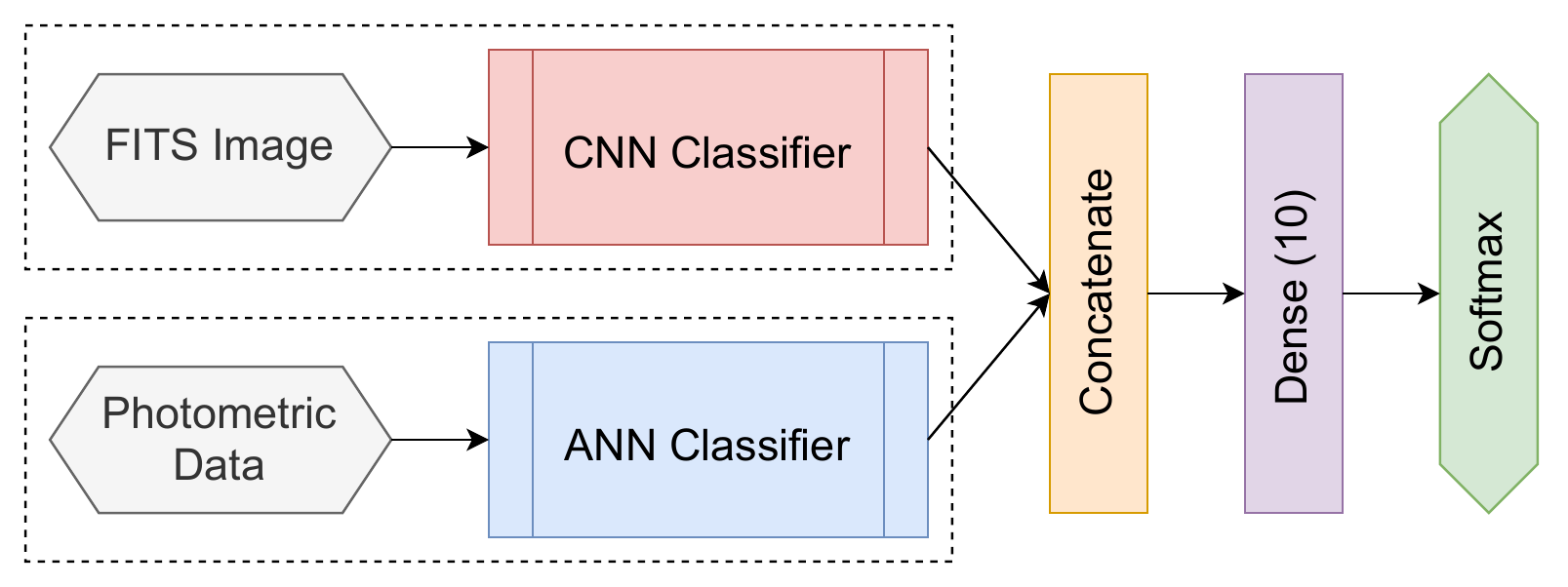}
\caption{A schematic representation of \texttt{MargNet}. Details of the full implementation are given in the Table \ref{tab:CNNmodel} and \ref{tab:annmodel}. Note that the outputs of the CNN classifier and the ANN classifier consist of 3 neurons each, which represent the separate predicted probabilities. These are then ``concatenated" together and fed forward to a fully connected dense layer of 10 neurons. These are then finally fed to 3 softmax activated neurons, which represent \texttt{MargNet}'s final probability predictions.}
\label{fig:ensemble_model}
\end{figure}

The CNN and ANN were trained with an early stopping condition that the training will stop if it shows no improvement in 30 consecutive epochs.
These previously-trained models are both stacked parallel to each other, in a stacking ensemble. We call our final network \texttt{MargNet}. We have optimised the CNN and ANN for their respective inputs and hence do not retrain \texttt{MargNet}. We take the objects in each class (both the images and photometric data) and give them as inputs for the two models, CNN and ANN, respectively. The outputs of these models then act as inputs for the ensemble (a schematic representation is shown in Figure \ref{fig:ensemble_model}). The expected outputs for both components are merged, and a statistical combination of them leads to the final output. We calculate the performance metrics for the final output. \texttt{MargNet} contains 103 neurons of its own and a total of 25,875,217 inclusive of CNN and ANN model trained neurons. 

\subsection{Evaluation Metrics}
\label{subsec:eval_metrics}

To measure classification performance, we evaluate three metrics - precision, recall and accuracy. Precision is the ratio of the true positives (TP) to the sum of true positives and false positives (FP), while recall ($R$) is the ratio of true positives to the sum of true positives and false negatives (FN). 

Thus, for a particular class $c$, the precision ($P_c$) and the recall ($R_c$) are given by:
\begin{equation}
    P_c = \frac{\text{TP}_c}{\text{TP}_c + \text{FP}_c},
    \\
    R_c = \frac{\text{TP}_c}{\text{TP}_c + \text{FN}_c},
\end{equation}
where the subscript $c$ denotes objects for that particular class only. We then calculate the total precision ($P$) and the total recall ($R$) by taking an average of the $P_c$ and $R_c$ across all classes\footnote{This is available on scikit-learn as the \texttt{macro} setting.}.

Intuitively, the precision for galaxies can be thought of as the classifier's ability to not misclassify a star/quasar as a galaxy, while
the recall for galaxies denotes the classifier's ability to identify all galaxies in the dataset.

Accuracy on the other hand, is the number of correctly classified samples divided by the total number of classified samples. This can be calculated for just a single class, or for all classes. Accuracy is thus given by,
\begin{equation}
    \text {Accuracy}=\frac{\text{TP}+\text{TN}}{\text{TP}+\text{TN}+\text{FP}+\text{FN}},
\end{equation}
where TN denotes the true negatives.

In this paper, all metrics refer to the total score across all classes, unless specified explicitly.

\section{Results and Discussion}\label{sec:results}
For each experiment, we work with two cases - star-galaxy classification and star-galaxy-quasar classification. We use the ensemble model, \texttt{MargNet}.

\subsection{Results: Experiment 1}
\label{subsec:exp1_results}
\begin{figure}%
    \centering
    \subfloat[Confusion matrix for star-galaxy classification: \texttt{MargNet} shows an accuracy of 98.23\% for galaxies and 98.03\% for stars individually. The star and galaxy mis-classifications are as low as \textasciitilde 1.87\% ]{{\includegraphics[width = 0.9\columnwidth]{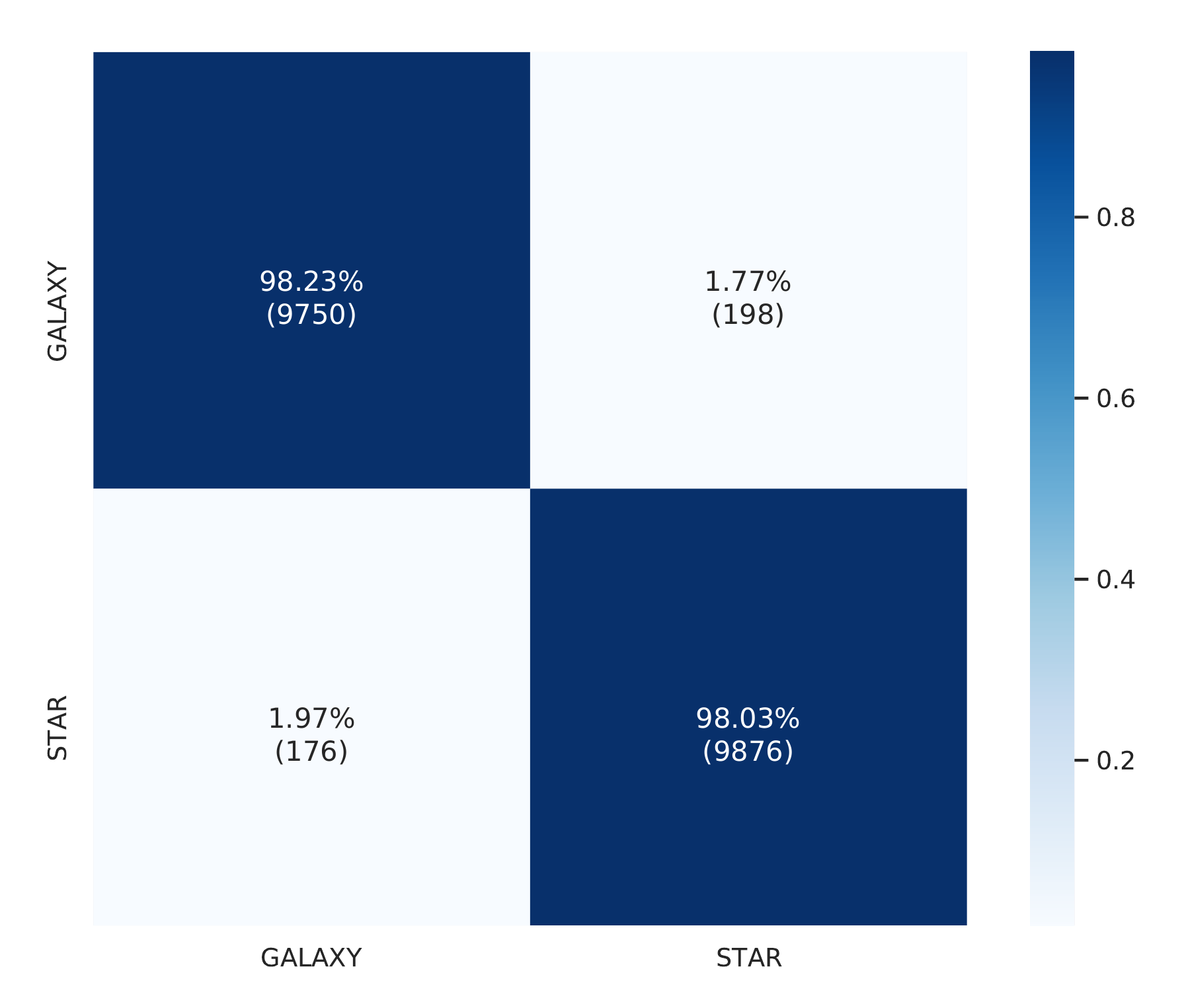} }\label{fig:ensemble_cm_sg}}%
    \qquad
    \subfloat[Confusion matrix for star-galaxy-quasar classification: \texttt{MargNet} shows an accuracy of 95.88\% for galaxies, 91.04\% for stars and 93.08\% for quasars individually. The average mis-classification rate is \textasciitilde 3.33\% ]{{\includegraphics[width = 0.9\columnwidth]{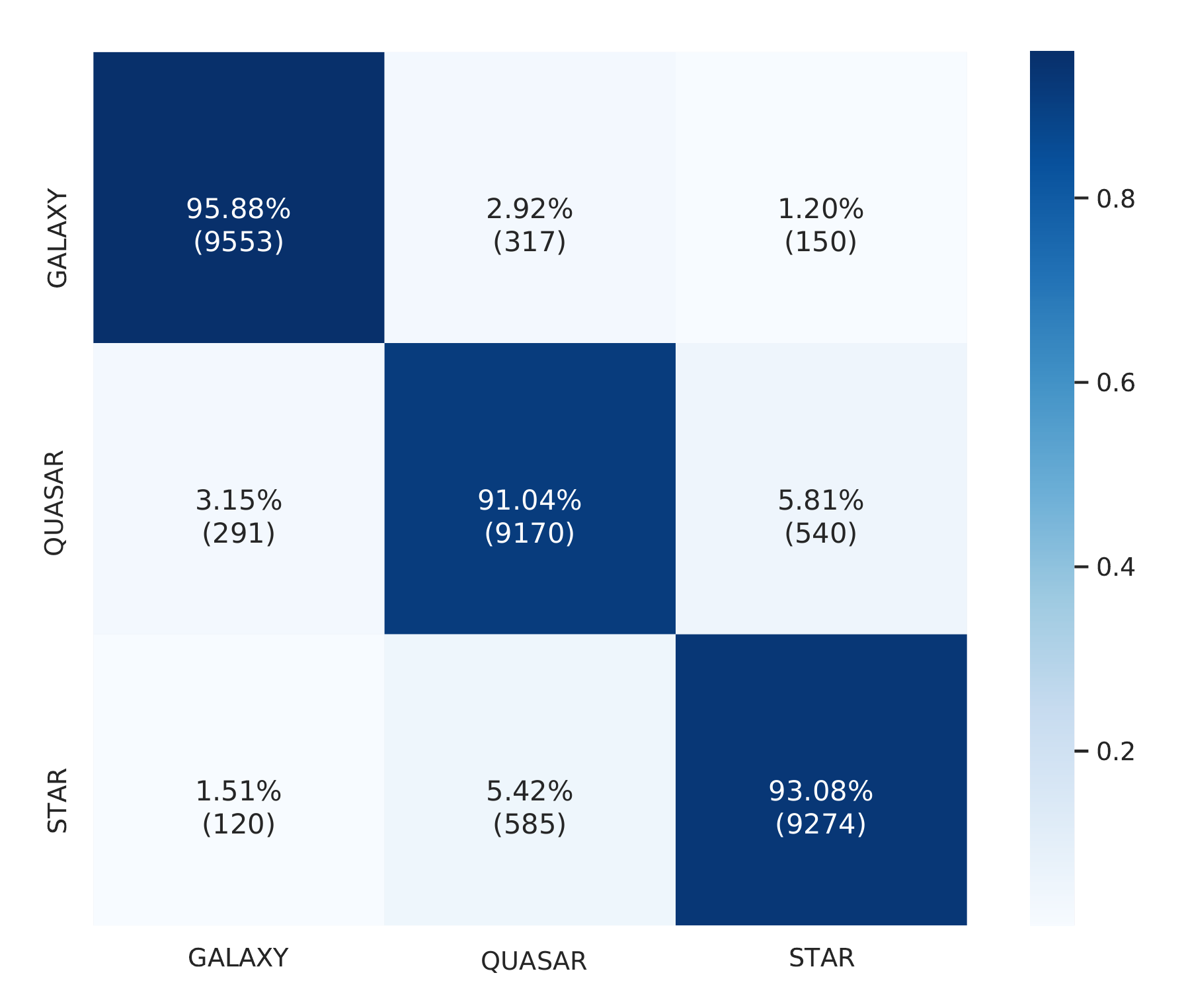} }\label{fig:ensemble_cm_sgq}}%
    \caption{Confusion Matrices of \texttt{MargNet} for Experiment 1: This shows the performance of \texttt{MargNet} for the both star-galaxy and the star-galaxy-quasar classification.}%
    \label{fig:ensemble_cm}
\end{figure}

In this experiment, all three sets - training, validation and test, are chosen from the Compact Source dataset.  We calculate the performance metrics for star-galaxy and star-galaxy-quasar classification. \texttt{MargNet} performs better than the ANN and CNN individually for both the classifications. (See Table \ref{tab:exp1_results}.)
For star-galaxy classification, we do not consider quasars because evaluating the performance for only stars and galaxies provides comparison with models used in other works. \texttt{MargNet} obtains an overall accuracy of 98.1 ± 0.1\% for the binary classification. We can also see from Figure \ref{fig:ensemble_cm_sg} that performance on both stars and galaxies is similar (98.23\% for galaxies and 98.03\% for stars), with a similar number of misclassifications for both (1.77\% for galaxies and 1.97\% for stars).

For star-galaxy-quasar classification, \texttt{MargNet} obtains an average accuracy across the three classes of of 93.3 ± 0.2\%. From the confusion matrices of figure \ref{fig:ensemble_cm_sgq}, it is evident that performance for stars, galaxies and quasars is different. Quasars have the lowest accuracy (91.04\%), and are misclassified as stars in \textasciitilde 5.8\% of the cases, and as galaxies in \textasciitilde 3.2\% of the cases. The more frequent identification of quasars as stars is understandable as both are point sources.  For the same reason, stars are misclassified as quasars in \textasciitilde 5.4\% of the cases.  Galaxies have the best individual accuracy among the three classes.  Galaxy-quasar misclassifications (galaxy-quasar: 2.92\% and quasar-galaxy: 3.15\% being the individual misclassification rates) are more common than galaxy-star misclassifications (galaxy-star: 1.20\% and star-galaxy: 1.51\%).
This is due to quasar host galaxy fuzz, which can be apparent for low-luminosity quasars, leading to their being misclassifed as a compact galaxy in some cases. Stars have no structure other than the PSF and are therefore less likely to be classified as a galaxy.

\begin{table*}
\centering
\resizebox{0.8\textwidth}{!}{%
\begin{tabular}{|l|c|c|ccc|} 
\hline
\multicolumn{1}{|c|}{Experiment} & Classification & Network & \multicolumn{1}{c|}{Accuracy (\%)} & \multicolumn{1}{c|}{Precision (\%)} & \multicolumn{1}{c|}{Recall (\%)} \\ 
\hline
\multicolumn{1}{|c|}{\multirow{6}{*}{Experiment 1}} & \multirow{3}{*}{1. Star-Galaxy} & ANN & 97.9 ± 0.1 & 97.8 ± 0.1 & 97.9 ± 0.1 \\
\multicolumn{1}{|c|}{} &  & CNN & 97.4 ± 0.1 & 97.4 ± 0.1 & 97.4 ± 0.1 \\ 
\cline{3-6}
\multicolumn{1}{|c|}{} &  & MargNet & 98.1 ± 0.1 & 98.1 ± 0.1 & 98.1 ± 0.1 \\ 
\cline{2-6}
\multicolumn{1}{|c|}{} & \multirow{3}{*}{2. Star-Galaxy-Quasar} & ANN & 93.0 ± 0.1 & 93.0 ± 0.1 & 93.0 ± 0.1 \\
\multicolumn{1}{|c|}{} &  & CNN & 91.6 ± 0.2 & 91.7 ± 0.2 & 91.6 ± 0.2 \\ 
\cline{3-6}
\multicolumn{1}{|c|}{} &  & MargNet & 93.3 ± 0.2 & 93.3 ± 0.2 & 93.3 ± 0.2 \\ 
\hline
\multirow{6}{*}{Experiment 2} & \multirow{3}{*}{1. Star-Galaxy} & ANN & 96.0 ± 0.1 & 96.0 ± 0.1 & 96.0 ± 0.1 \\
 &  & CNN & 95.2 ± 0.1 & 95.2 ± 0.1 & 95.2 ± 0.1 \\ 
\cline{3-6}
 &  & MargNet & 96.9 ± 0.1 & 96.9 ± 0.1 & 96.9 ± 0.1 \\ 
\cline{2-6}
 & \multirow{3}{*}{2. Star-Galaxy-Quasar} & ANN & 86.0 ± 0.1 & 86.1 ± 0.1 & 86.1 ± 0.1 \\
 &  & CNN & 84.2 ± 0.2 & 84.5 ± 0.2 & 84.2 ± 0.2 \\ 
\cline{3-6}
 &  & MargNet & 86.7 ± 0.2 & 86.8 ± 0.2 & 86.7 ± 0.2 \\ 
\hline
\multirow{6}{*}{Experiment 3} & \multirow{3}{*}{1. Star-Galaxy} & ANN & 91.6 ± 0.1 & 92.4 ± 0.1 & 91.6 ± 0.1 \\
 &  & CNN & 89.3 ± 0.1 & 90.4 ± 0.1 & 89.3 ± 0.1 \\ 
\cline{3-6}
 &  & MargNet & 92.0 ± 0.1 & 92.7 ± 0.1 & 92.0 ± 0.1 \\ 
\cline{2-6}
 & \multirow{3}{*}{2. Star-Galaxy-Quasar} & ANN & 73.1 ± 0.1 & 76.3 ± 0.1 & 73.1 ± 0.1 \\
 &  & CNN & 69.4 ± 0.2 & 73.9 ± 0.2 & 69.4 ± 0.2 \\ 
\cline{3-6}
 &  & MargNet & 73.4 ± 0.2 & 76.5 ± 0.2 & 73.4 ± 0.2 \\
\hline
\end{tabular}%
}
\caption{\texttt{MargNet} performance metrics: This table shows the performance of our models ANN, CNN and the ensemble (\texttt{MargNet}) considering our compactness criterion mentioned in the Section \ref{subsec:compact-faint}, and faintness criteria based on the r-band magnitude. We show the performance metrics for all the three experiments which were carried out. Details of these experiments are mentioned in Section \ref{sec:intro}. We see that \texttt{MargNet} outperforms the individual models and gives good accuracy, and balanced precision and recall values for the star-galaxy and the star-galaxy-quasar problem for all the three experiments. The detailed confusion matrices for \texttt{MargNet} for both the problems can be seen in Figure \ref{fig:ensemble_cm}.}
\label{tab:exp1_results}
\end{table*}

\subsection{Results: Experiment 2}
\label{subsec:exp2_results}
\begin{figure*}%
    \centering
    \subfloat[Star-galaxy (SG) and star--galaxy-quasar (SGQ) classification for Experiment 2: For SG, we see accuracy sharply decreases at $r = 21$ and an unexpected rise for $r > 22$.For SGQ, similar trend as from SG case. In addition to SG, for quasar recall and quasar precision a rise follows a drop can be seen. At $r > 22$ there is no rise seen in galaxy precision. For further details refer to Section \ref{subsec:exp2_results} ]{{\includegraphics[width = \columnwidth]{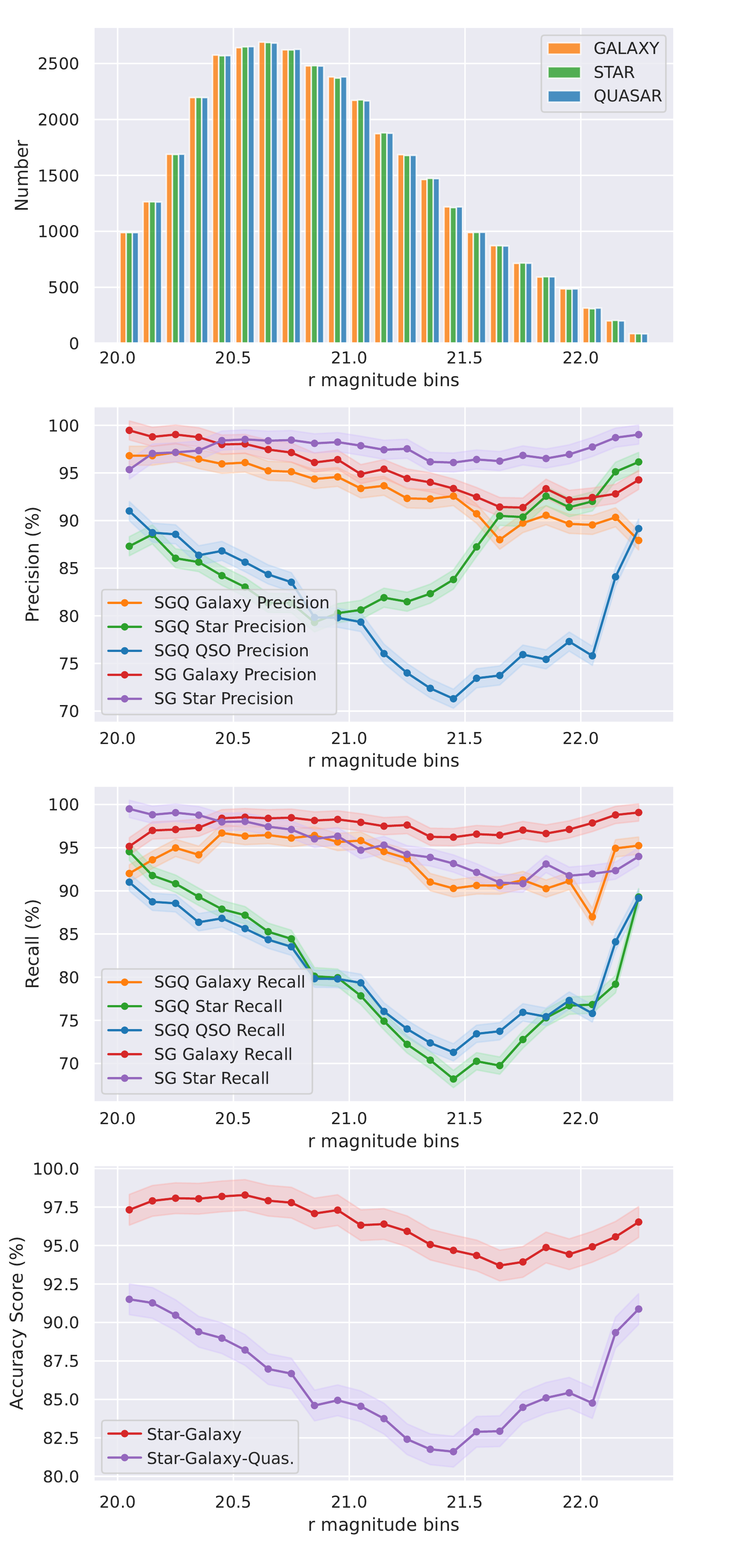} }\label{fig:exp2_results}}%
    \qquad
    \subfloat[SG and SGQ classification for Experiment 3: For SG, the trends are similar to Experiment 2. For SGQ, the trends are similar to Experiment 2 (Figure \ref{fig:exp2_results}). Here, for quasar recall and quasar precision a rise follows a drop can be seen. At $r > 22$ there is no rise seen in galaxy precision as compared to the Figure\ref{fig:exp2_results}. ]{{\includegraphics[width = \columnwidth]{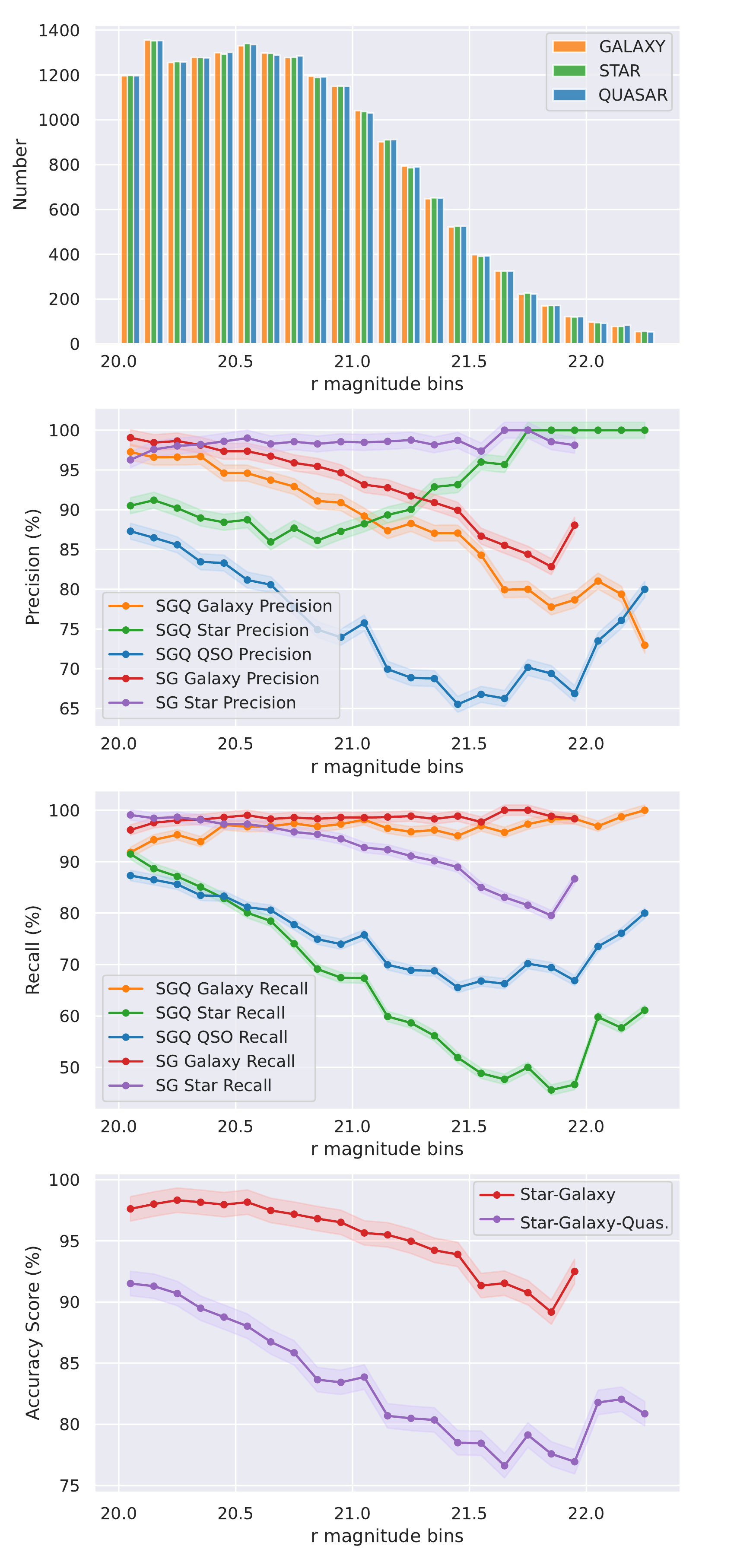} }\label{fig:exp3_results}}%
    \caption{Performance for Experiment 2: For the faint and compact dataset $c<0.5; r > 20$. The data is divided into bins of  width 0.1 magnitude with the final range in $20 < r < 22.6$. Precision, recall and accuracy metrics are evaluated for star-galaxy and star-galaxy-quasar problem, shown in Figures \ref{fig:exp2_results}. Performance for Experiment 3: Here training is done on the compact dataset $c<0.5$, but testing is done on the faint and compact dataset $c<0.5; r > 20$. This data is divided into bins of 0.1 magnitudes with the final range in $20 < r < 22.6$. Precision, recall and accuracy metrics are evaluated for star-galaxy and star-galaxy-quasar problem, shown in Figure \ref{fig:exp3_results}.}%
    \label{fig:exp_results}
\end{figure*}

In this experiment, all three sets - training, validation and test, are chosen from the Faint and Compact Source dataset. We evaluate the statistics as a function of the magnitude to study the performance of \texttt{MargNet} at fainter magnitudes (Table \ref{tab:exp1_results}). For this, we divide the test set into bins of width 0.1 magnitude in the range $20<r<22.6$.  Each bin in this range has at least 50 objects per class, which is a large enough sample size for the analysis.   We evaluate different metrics for each bin and plot those in Figure \ref{fig:exp2_results}.
We observe a decrease in performance as compared to Experiment 1 because the fainter an object is, the lower its is Signal-to-Noise Ratio (SNR). This noise then propagates as an error in the object's photometric measurements and calculated features, making classification difficult.

For the star-galaxy binary classification problem, the a trend we see Figure \ref{fig:exp2_results} is that classification accuracy steadily decreases upto $r=21.5$. We expect this, as the SNR in the photometric measurements reduces at fainter magnitudes. We note that there is an unexpected rise in accuracy for $r>21.5$. We observe a similar trend for galaxy precision and star recall - a rise follows a drop. However, star precision and galaxy recall maintain a more or less constant value. This implies that, between $21<r<22$, the probability of a source predicted as a star being a star is very high. On the other hand, the probability of a source predicted as a galaxy not being a galaxy is very low.

In \ref{fig:exp2_results}, for the tri-class classification problem of stars-galaxies-quasars, we see a trend similar to the one seen in Experiment 1 (Figure \ref{fig:exp_results}). For classification accuracy, star recall, quasar recall and quasar precision, a rise follows a drop, similar to the one we saw earlier in Section \ref{subsec:exp2_results}. 
Even the star precision and galaxy recall values are almost perfect and constant as earlier.
However, the one that behaves differently is galaxy precision. At $r>22$, unlike the star-galaxy case, no rise is seen. Instead, it just plateaus.

\subsection{Results: Experiment 3}
\label{subsec:exp3_results}
In this experiment, the training and validation set is chosen from the Compact Source dataset, while the test set is chosen from the Faint and Compact Source dataset. Here, we present the performance at various magnitudes, as in Experiment 2, dividing the test set for Experiment 3 into bins of width 0.1 magnitudes. We plot the different metrics for each bin in Figure \ref{fig:exp3_results}.
We expect a further decrease in performance when compared to Experiment 2, because the test set is not representative of the training set here. The training set consists of only compact sources, where the magnitude of the set covers a wide range from $14 - 23$ (Figure \ref{fig:distribution}). However, the test set consists of only faint and compact sources in the range $20 - 22.6$.

The trend we see from Figure \ref{fig:exp3_results} is similar to Experiment 2 (Figure \ref{fig:exp2_results}), albeit with a decreased performance overall, as expected.
We see that even though the test set is not representative of the training set, it still performs fairly well for star-galaxy classification.

However, when considering quasars in the classification. The star-galaxy-quasar classification shows (Figure \ref{fig:exp3_results}) that the performance reduces even before $r=21$. Furthermore, the rise in performance at higher magnitudes is almost insignificant, which is unlike the star-galaxy case from Experiment 3 (Section \ref{subsec:exp3_results}) and Experiment 2 (Section \ref{subsec:exp2_results}).

\subsection{Discussion}
The paper by \citet{sebok1986angular} was an early foray into object classification and finding the magnitude limits for the confident separation of stars and galaxies. It aimed to find a magnitude to which a decision envelope can be drawn between stars and galaxies. This study for 1,30,000 objects found that objects fainter than $r = 19.6$ are not separated. The classification envelope crosses the decision boundary of $\phi$ = 1. This criterion is a design of the classification and pattern recognition system used by the author, called a phi classifier, which works with linear data. \texttt{MargNet}, on the other hand, separates galaxies and stars with high completeness until $r=21.5$ with a drop at fainter magnitudes. While other techniques after \citet{sebok1986angular} successfully classified faint objects with r>19.6, we demonstrate that \texttt{MargNet} yields better performance than some well-known results as mentioned below.

In Figure \ref{fig:exp_results}, we observe that the accuracy starts rising up beyond the 21.5 magnitude in the r-band for both the star-galaxy and star-galaxy-quasar problems. We do not expect such a trend, because fainter galaxies are more difficult to classify, and should thus lead to a low accuracy. 
A possible reason for this could be that our training dataset consists of an equal number of stars, galaxies and quasars but their distribution for various r-band magnitudes is different - there are 2831 galaxies, 190 stars and 660 quasars in our training set which satisfy the condition $r > 21.5$. 

On the other hand, during testing, we use an equal number of all three objects across each magnitude bin, to ensure that the performance statistics for a particular bin are not influenced because of class imbalance. We believe that due to this large bias during training, the model's behaviour changes. Galaxies are less likely to be classified as another object, relatively reducing its number of false negatives. This effectively reduces the number of false positives for stars and quasars, and likely contributes to the increased precision.

To compare \texttt{MargNet}'s results with results in \citet{kimStarGalaxyClassification2017}, we plot the precision (purity) and recall (completeness)\footnote{Purity and completeness are conventional terms used in earlier astronomy literature, and mean the same as precision and recall, respectively.} again but impose additional conditions. \footnote{Here we note that the training set of \citet{kimStarGalaxyClassification2017} is different and the comparison is not easy as we cannot comment on whether the different results are  due to the use of \texttt{MargNet} or due to different sky cuts for the data.}For galaxies, we plot the precision while maintaining constant recall ($c_g = 0.96$). For stars, we plot the recall while maintaining constant precision ($p_g = 0.97$). We observe that \texttt{MargNet} consistently outperforms \citet{kimStarGalaxyClassification2017} when comparing galaxies in the faint regime, $r>21$, as per our aim. We also notice a sharp minimum between $20<r<20.25$ for galaxy precision, which is unexplained. For stars, \texttt{MargNet}'s recall exceeds \citet{kimStarGalaxyClassification2017}, after which it sharply drops at around $r=21.5$. We would also like to clarify some gaps in both plots: we could not achieve the additional criterion ($c_g=0.96$ for galaxies and $p_s=0.97$ for stars) at these points.

\begin{figure}
    \centering
    \includegraphics[width=\columnwidth]{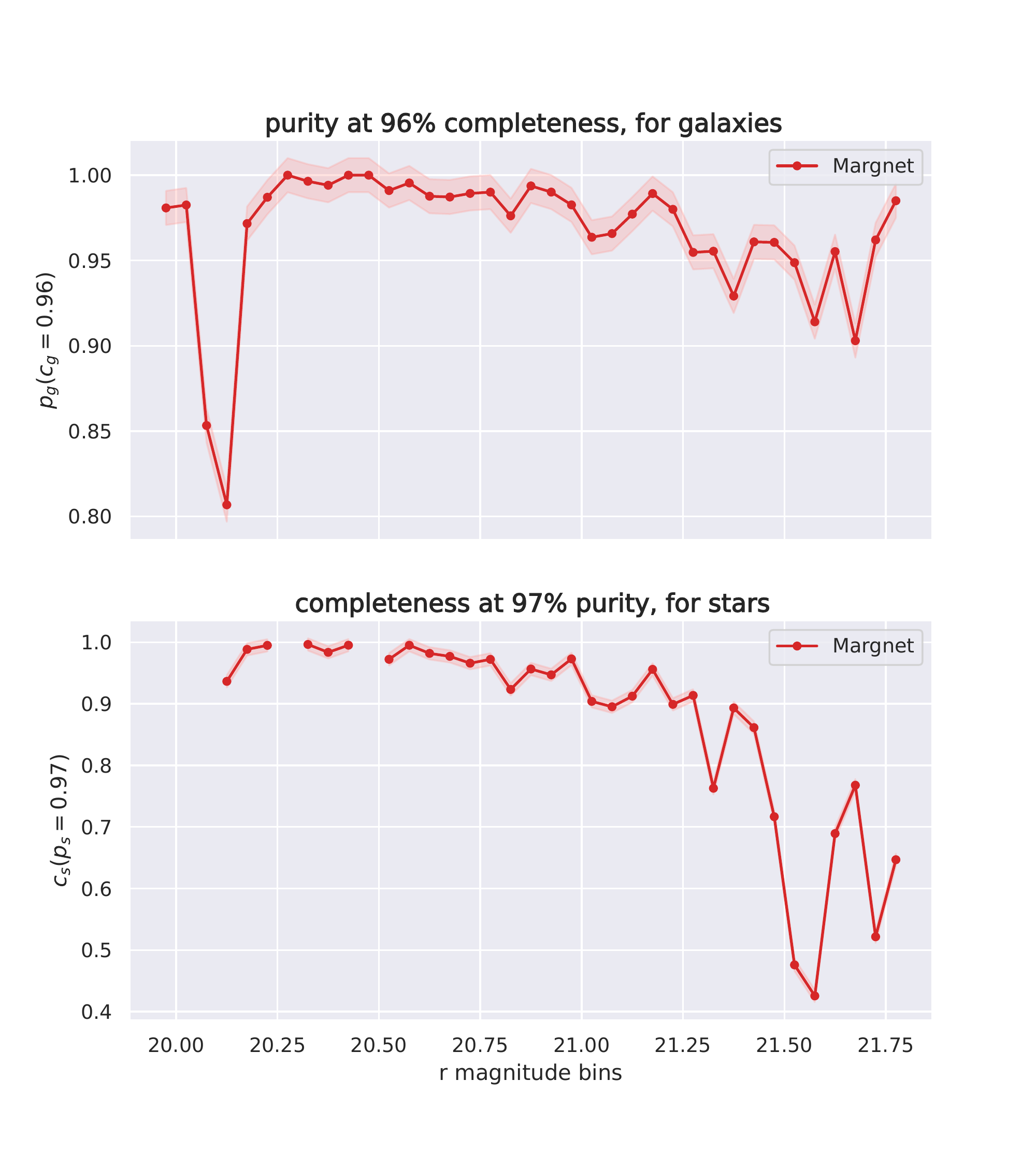}
    \caption{Galaxy purity (precision) and star completeness (recall) with additional conditions $(c_g = 0.96; p_s = 0.97)$ as a function of the r-band magnitude. Note that we use i magnitude in the x axis to compared it with Figure 8 from \citet{kimStarGalaxyClassification2017}.}
    \label{fig:comparison_kimbrunner}
\end{figure}

\citet{soumagnacStarGalaxySeparation2015} considered star-galaxy separation in the context of their work on Large Scale Structure and Weak Lensing using the Dark Energy Survey (DES). Comparing our results with them we find that \texttt{MargNet} performs better in the r = 20 to 20.5 regime. However, their model performs better for r = 20.5 to 23 magnitudes as the survey that they use is a deep and wide survey, with  a large number of objects at the faint magnitudes for the training. 

We can also compare our results with star-galaxy-quasar classification carried for other datasets \citep[e.g.:,][]{Xiaoqing2020ClassificationOS,nakazono}. We note  that such comparison is not straightforward for data  from different surveys with different data selection criteria. \citet{Xiaoqing2020ClassificationOS} used different machine learning approaches to classify stars, galaxies and quasars from the Large Sky Area Multi-Object Fiber Spectroscopic telescope \citep[LAMOST;][]{lamost}, and obtained their best accuracy (97.3 \%) using a Random Forest Classifier. \citet{nakazono} compared the performance of Support Vector Machines and Random Forests for the classification of stars/galaxies/quasars in the Southern Photometric Local Universe Survey \citep[S-PLUS;][]{splus}, supplemented by data from the Wide-field Infrared Survey Explorer \citep[WISE;][]{wise2010}. They achieved an averaged F-score of $96.85 \pm 0.05$ by using a Random Forest with 12 S-PLUS bands + 2 WISE bands.

\begin{figure}
    \centering
    \includegraphics[width=\columnwidth]{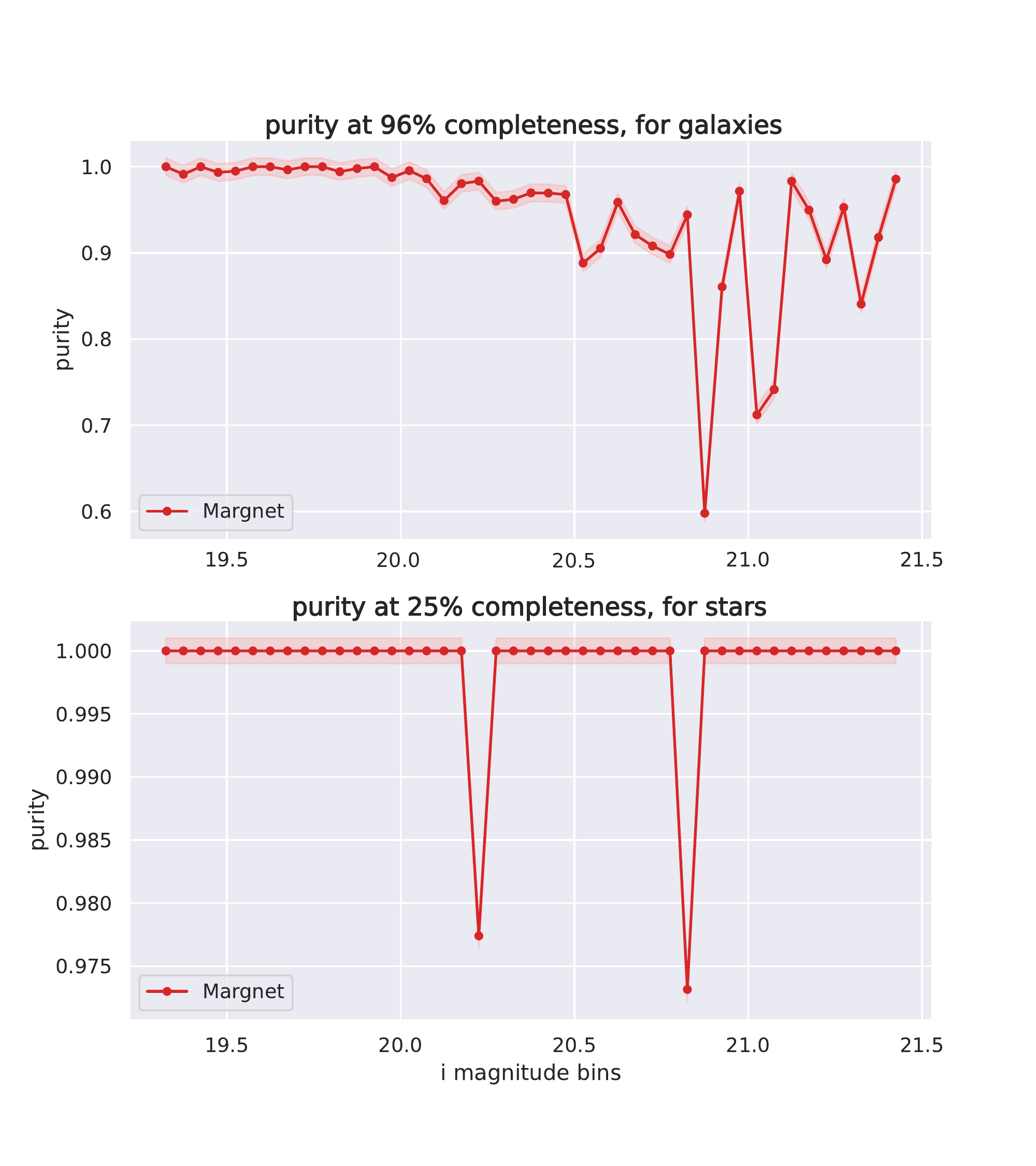}
    \caption{Galaxy purity (precision) and star purity with additional conditions on completness (recall) $(c_g = 0.96; c_s = 0.25)$ as a function of the r-band magnitude. This plot can be compared with Figure 10 from \citet{soumagnacStarGalaxySeparation2015}}
    \label{fig:comparison_soumagnac}
\end{figure}

\section{Conclusion \& Future Scope}
\label{sec:conclusion}
We have presented a multiple neural network model named \texttt{MargNet} for classifying stars, galaxies and quasars in the SDSS photometric images. For the SDSS dataset, \texttt{MargNet} can provide a classification that is more accurate than previous works involving neural networks for compact and faint galaxies. The major advantage of \texttt{MargNet} is that useful features are learned automatically from images, while traditional machine learning algorithms require feature engineering as a separate process to produce accurate classifications. A part of \texttt{MargNet} is built by inception net modules. Inception Nets have recently achieved promising results in many image classification tasks and have been quickly and widely adopted by the computer vision community \citep{xiong2018ai}. Usual networks show low accuracy when the sources become increasingly faint and compact,  but \texttt{MargNet} shows promising results for such sources, with an overall accuracy of $\sim 93.3\%$. For these faint and compact sources, the CNN and ANN individually perform with \textasciitilde 91.6\% and \textasciitilde 93.0\% accuracy, respectively. This denotes improvement in accuracy when both parts are run together. \texttt{MargNet} should be useful for future surveys like LSST which will capture many faint and compact sources. \texttt{MargNet} should also be useful for other surveys like GAIA and ZTF using crossmatching to SDSS and transfer learning.  Although we will need photometric features to use the ensemble model, these features need not be identical to those from the SDSS. This could be done by altering the ANN network to accommodate other available features. Even in the case no such photometric features are obtainable, the image-based CNN component of \texttt{MargNet} could be used individually. The methodology behind \texttt{MargNet} will thus be of use with its accuracy in classifying objects as seen by such deeper astronomical surveys.

\section*{Acknowledgements}

S. Chaini acknowledges the computing resources provided by the Computer Centre, IISER Bhopal. M. Vivek acknowledges support from DST-SERB in the form of the core research grant (CRG/2020/1657).
A. Bagul acknowledges support from Ms. Namita Yadav in the form of visualisations of models using figures.

We are grateful to the Sloan Digital Sky Survey for making their data free and open source.
Funding for SDSS IV has been provided by the Alfred P. Sloan Foundation, the U.S. Department of Energy Office of Science, and the Participating Institutions. 
SDSS-IV acknowledges support and resources from the Center for High Performance Computing  at the University of Utah. The SDSS website is www.sdss.org.
SDSS-IV is managed by the Astrophysical Research Consortium for the Participating Institutions of the SDSS Collaboration including the Brazilian Participation Group, 
the Carnegie Institution for Science, Carnegie Mellon University, Center for 
Astrophysics | Harvard \& Smithsonian, the Chilean Participation Group, the French Participation Group, Instituto de Astrof\'isica de Canarias, The Johns Hopkins 
University, Kavli Institute for the Physics and Mathematics of the Universe (IPMU) / University of Tokyo, the Korean Participation Group, Lawrence Berkeley National Laboratory, 
Leibniz Institut f\"ur Astrophysik Potsdam (AIP),  Max-Planck-Institut 
f\"ur Astronomie (MPIA Heidelberg), Max-Planck-Institut f\"ur Astrophysik (MPA Garching), Max-Planck-Institut f\"ur Extraterrestrische Physik (MPE), National Astronomical Observatories of China, New Mexico State University, New York University, University of 
Notre Dame, Observat\'ario Nacional / MCTI, The Ohio State University, Pennsylvania State 
University, Shanghai Astronomical Observatory, United Kingdom Participation Group, 
Universidad Nacional Aut\'onoma de M\'exico, University of Arizona, University of Colorado Boulder, University of Oxford, University of Portsmouth, University of Utah, University of Virginia, University of Washington, University of Wisconsin, Vanderbilt University, and Yale University.

\emph{Software:} Keras \citep{chollet2015keras}, TensorFlow \citep{tensorflow},  NumPy \citep{numpy,numpy2}, Jupyter \citep{jupyter}, Matplotlib \citep{matplotlib}, Seaborn \citep{seaborn} scikit-learn \citep{scikit-learn}, SciPy \citep{scipy}, Pandas \citep{pandas,pandas2}, tqdm \citep{tqdm}, KerasTuner \citep{kerastuner} and Python3 \citep{python3}.

%%%%%%%%%%%%%%%%%%%%%%%%%%%%%%%%%%%%%%%%%%%%%%%%%%
\section*{Data Availability}

All our code used for this paper has been made open-source on GitHub at the URL: \url{https://github.com/sidchaini/MargNet}. We downloaded table data from SDSS Casjobs DR16\footnote{\url{https://skyserver.sdss.org/casjobs/}}, using the query listed in Appendix \ref{app:sdssquery}. We downloaded image data from SDSS\footnote{\url{https://data.sdss.org/sas/dr16/eboss/photoObj/}} using a Python script which is available on our GitHub repository. Our complete preprocessed dataset and our trained models are available on Zenodo\footnote{\url{https://doi.org/10.5281/zenodo.6659435}} \citep{margnet_dataset}.

%%%%%%%%%%%%%%%%%%%% REFERENCES %%%%%%%%%%%%%%%%%%

% The best way to enter references is to use BibTeX:

\bibliographystyle{mnras}
\bibliography{references} % if your bibtex file is called example.bib

% Alternatively you could enter them by hand, like this:
% This method is tedious and prone to error if you have lots of references
%\begin{thebibliography}{99}
%\bibitem[\protect\citeauthoryear{Author}{2012}]{Author2012}
%Author A.~N., 2013, Journal of Improbable Astronomy, 1, 1
%\bibitem[\protect\citeauthoryear{Others}{2013}]{Others2013}
%Others S., 2012, Journal of Interesting Stuff, 17, 198
%\end{thebibliography}

%%%%%%%%%%%%%%%%%%%%%%%%%%%%%%%%%%%%%%%%%%%%%%%%%%

%%%%%%%%%%%%%%%%% APPENDICES %%%%%%%%%%%%%%%%%%%%%

\appendix

\section{SDSS SQL Query} \label{app:sdssquery}

We used the following SQL Query on CasJobs to obtain all
photometric features used in our work.
\begin{verbatim}
SELECT TOP 50000 p.objid, s.specobjid, 
s.class,p.run, p.rerun, p.camcol, p.field,
p.ra, p.dec, s.z as redshift,
p.dered_u, p.deVRad_u, p.psffwhm_u, p.extinction_u,
p.dered_g, p.deVRad_g, p.psffwhm_g, p.extinction_g,
p.dered_r, p.deVRad_r, p.psffwhm_r, p.extinction_r,
p.dered_i, p.deVRad_i, p.psffwhm_i, p.extinction_i,
p.dered_z, p.deVRad_z, p.psffwhm_z, p.extinction_z,
p.u - p.g AS u_g,
p.g - p.r AS g_r,
p.r - p.i AS r_i,
p.i - p.z AS i_z
FROM SpecObj as s
JOIN PhotoObj AS p ON p.objid = s.bestObjID
WHERE s.z > 0 AND s.zErr<0.1
AND s.class = 'STAR' -- Put STAR/GALAXY/QSO
AND s.bestObjID = s.fluxObjID
--For faint objects
AND (p.u+p.g+p.r+p.i+p.z)/5 > 20
AND p.psffwhm_u !=0 AND p.psffwhm_g !=0 
AND p.psffwhm_r !=0 
AND p.psffwhm_i !=0 AND p.psffwhm_z !=0
--For compact objects
AND ((p.deVRad_u / p.psffwhm_u) + 
    (p.deVRad_g / p.psffwhm_g) +
    (p.deVRad_r / p.psffwhm_r) + 
    (p.deVRad_i / p.psffwhm_i) + 
    (p.deVRad_z / p.psffwhm_z)) / 5 < 0.5
AND ((p.err_u/p.u) + (p.err_g/p.g) +
    (p.err_r/p.r) + (p.err_i/p.i) +
    (p.err_z/p.z)) / 5 < 0.1
AND p.deVRad_u != 0 AND p.deVRad_g !=0
AND p.deVRad_r !=0 
AND p.deVRad_i !=0 AND p.deVRad_z !=0
-- Quality flags
AND s.zWarning=0 AND p.clean=1 AND p.mode = 1 
AND s.sciencePrimary = 1
AND s.targetType = 'SCIENCE' 
AND p.insideMask=0
AND (p.flags_r & 0x20) = 0
AND (p.flags_r & 0x80000) = 0
AND (p.flags_r & 0x800000000000) = 0
AND (p.flags_r & 0x10000000000) = 0
AND ((p.flags_r & 0x100000000000) = 0 
OR (p.flags_r & 0x1000) = 0)  
AND (p.flags_r & 0x40000) = 0
AND (p.flags_r & 0x80) = 0
AND (p.flags_r & 0x4) = 0
AND ((p.flags_r & 0x10000000) != 0)
AND ((p.flags_r & 0x8100000c00a0) = 0)
AND (((p.flags_r & 0x400000000000) = 0) 
AND (p.psfmagerr_r <= 0.2))
AND (((p.flags_r & 0x100000000000) = 0) 
AND (p.flags_r & 0x1000) = 0);
\end{verbatim} 

%%%%%%%%%%%%%%%%%%%%%%%%%%%%%%%%%%%%%%%%%%%%%%%%%%

% Don't change these lines
\bsp	% typesetting comment
\label{lastpage}
\end{document}